\documentclass[twocolumn, prd,showpacs,showkeys]{revtex4}      

\usepackage{amsfonts}
\usepackage{amsmath}
\usepackage{amssymb}
\usepackage{bbm}
\usepackage{bm}
\usepackage[utf8]{inputenc}
\usepackage{graphicx}%
\usepackage{tensor} 	
\usepackage{slashed}	
\usepackage[centertableaux]{ytableau}
\usepackage{hyperref}
\usepackage{natbib}
\usepackage{longtable}

\newcommand{\ket}[1]{\ensuremath{\left |#1 \right\rangle}}
\newcommand{\poi}{Poincar\'{e}}
\newcommand{\cas}[1]{\mathcal{C}_{#1} }

\newcommand{\so}{\ensuremath{\mathfrak{so}}}

\newcommand{\su}{\ensuremath{\mathfrak{su}}}
\renewcommand{\vec}[1]{\ensuremath{\mathbf{#1}}}

\begin{document}
\title{Covariant basis induced by parity for the $(j,0)\oplus (0,j)$ representation}
\author{Selim G\'{o}mez-\'{A}vila and M. Napsuciale}
\affiliation{Departamento de F\'{\i}sica, Universidad de Guanajuato, Campus Le\'{o}n; Lomas del Bosque 103,
Fraccionamiento Lomas del Campestre, 37150, Le\'{o}n, Guanajuato, M\'{e}xico.}

\begin{abstract}
In this work, we build a covariant basis for operators acting on the $(j,0)\oplus(0,j)$ Lorentz group representations. 
The construction is based on an analysis of the covariant properties of the parity operator, which for these representations 
transforms as the completely temporal component of a symmetrical tensor of rank $2j$. The covariant properties of parity 
involve the Jordan algebra of anti commutators of the Lorentz group generators which unlike the Lie algebra is not 
universal. We make the construction explicit for $j=1/2,1$ and $3/2$, reproducing well-known results for the $j=1/2$ case.
We provide an algorithm for the corresponding calculations for arbitrary $j$. This covariant basis provides an inventory 
of all the possible interaction terms for gauge and non-gauge theories of fields for these representations. In particular, 
it supplies a single second rank antisymmetric structure, which in the \poi\ projector formalism implies a single Pauli 
term arising from gauge interactions and a single (free) parameter $g$, the gyromagnetic factor. This simple structure predicts 
that for an elementary particle in this formalism all multipole moments, $Q^{l}_{E}$ and $Q^{l}_{M}$, are dictated by the 
complete algebraic structure of the Lorentz generators and the value of $g$. We explicitly calculate the multipole moments, 
for arbitrary $j$ up to $l=8$. Comparing with results in the literature we find that only the electric charge and magnetic 
moment of a spin $j$ particle are independent of the Lorentz representation under which it transforms, all higher multipoles 
being representation dependent. Finally we show that the propagation of the corresponding spin $j$ waves in a electromagnetic 
background is causal. 
\end{abstract}
\maketitle 
\section{Introduction}

The Standard Model of particle physics has three ingredients: a base spacetime, whose symmetries allow us to classify 
the asymptotically free states, a gauge group, which fixes the number and properties of gauge particles, and a particular 
spectrum of matter particles, whose number and interactions are largely unfixed \cite{McCabe:1087014}. Supersymmetry and 
extra-dimensional models represent attempts to modify our conventional understanding of the first ingredient, while 
extensions of the gauge group, including grand unified theories are extrapolations of the second one which also enlarge the 
content of spin 1/2 matter fields. The recent discovery of a particle with a mass around $126~ GeV$ at the LHC 
\cite{Chatrchyan201230, Aad20121} as expected for the Higgs of the standard model seems discouraging for both approaches 
at least in the short run\cite{Shifman:2012na}, and therefore it may prove valuable to focus on different avenues for going 
beyond the standard model (SM). 

In this concern, it is worth to remark that the SM makes use of only a few representations of the homogeneous Lorentz group which 
are identified with the three basic types of fields entering this construction. The $(0,0)$ representation for the Higgs field, 
the $(1/2, 1/2)$ representation for the gauge fields and the $(1/2,0)$ and $(0,1/2)$ representation for the matter fields.  
Certainly, so far we have no general principle restricting the spin content of these building blocks and some theories for
physics beyond the standard models like supersymmetric models modify this correlation and consider Lorentz 
representations with high spin content, the spinor-vector representation $[(1/2,0)\oplus(0,1/2)]\otimes (1/2,1/2)$, which 
enter as  gauge fields.    

High-spin fields naturally appear also in phenomenology (e.g., the hadronic contribution to the leptonic $g-2$) and in 
beyond the SM model building (e.g., supergravity, strings). An explanation of the family structure of the SM involving 
compositeness is also likely to imply the existence of some high-spin states. However, the description of high ($j>1$) spin 
elementary systems is a long-standing problem in quantum field theory. 
 
Conventional constructions for high spin fields (like Dirac-Fierz-Pauli \cite{Dirac:1936tg,Fierz:1939ix} or Rarita-Schwinger 
\cite{Rarita:1941mf}) are plagued by well-known problems, which can be traced to ambiguities in the selection and propagation 
of the desired degrees of freedom. These ambiguities can give rise to inconsistencies like superluminal propagation and other 
non localities, or the appearance of ghosts 
\cite{Velo:1969bt,Velo:1970ur,Velo:1972rt, Johnson:1960vt, Capri:1975si, Deser:2003ew, Porrati:2008gv}.  

Quantum fields which satisfy the cluster decomposition principle \cite{Weinberg:1995mt} are built as induced representations 
of the semi-direct product $\mathfrak{t}_{4} \rtimes \so{(1,3)}$ on representations of the Lorentz algebra. This often means 
that we use fields with redundant or unwanted degrees of freedom. We can generically understand the Velo-Zwanziger problems 
as a failure to discard these degrees of freedom in the presence of interactions.  

We can identify three classes of high-spin constructions. There are those in the spirit of Dirac-Fierz-Pauli, where either 
constrictions or auxiliary fields are used to remove unwanted degrees of freedom in order to exclusively propagate an 
irreducible representation of the \poi\ group 
\cite{Singh:1974qz, Singh:1974rc, Dirac:1936tg, Fierz:1939ix, Rarita:1941mf, Chang:1967zzc}. Then, there are constructions which 
renounce to fixing a single spin and mass, and contain several fields, as per example in Bhabha or Kemmer-Duffin-Petiau 
\cite{Bhabha:1945zz, Bargmann:1948ck, PhysRev.102.568, Green:1977nx} theories. Finally, we have the Joos-Weinberg formalism where 
no extra spin degrees of freedom are introduced\cite{Joos:1962qq, Weinberg:1964cn, Weinberg:1964ev, Tung:1967zz}
(see \cite{Esposito:2011wf} for a historical account.) 

Although the properties of free fields are only related to the \poi\ quantum numbers, regardless of the Lorentz representations 
on which the fields are constructed,  this is no longer true for interacting fields. Two quantum fields 
with spin $j$ quanta but defined in different Lorentz representations may share the same asymptotically free properties, but differ 
in their interaction properties, like their electromagnetic moments \cite{DelgadoAcosta:2012yc}. A basic element in the 
elucidation of the different possibilities to describe interactions of high spin systems is the classification of all possible 
operators acting in the corresponding Lorentz representation space, which in turn requires the construction of a covariant basis 
for these operators.

In a recent series of works \cite{Napsuciale:2006wr, Napsuciale:2007ry, AngelesMartinez:2011nt, VaqueraAraujo:2012qa}, a proposal 
for the description of arbitrary spin particles was detailed, based on the projection onto subspaces of the \poi\ group for 
fields with definite quantum numbers transforming in a given representation of the Lorentz group. A key ingredient in the formalism 
is the construction of the most general space-time antisymmetric operator in the corresponding space. In the general case, the 
construction of this tensor requires to classify all operators acting on the chosen representation space in a covariant manner. 
In this work we solve this problem for particles of arbitrary spin $j$ transforming in the $(j, 0) \oplus (0, j)$ 
representation of the Lorentz group.  We find a covariant basis for all operators with internal indexes in these representations. 
We conclude that there is a single independent antisymmetric operator which gives rise to a unique Pauli-type interaction term  
for all representations $(j, 0) \oplus (0, j)$ and work out the consequences for the multipole electromagnetic moments and the 
propagation of spin $j$ waves in an electromagnetic background. 

Although solving this problem is the main motivation for this work, the scope of the obtained results is beyond this framework 
since our covariant basis can be helpful in general for the construction of models using the $(j, 0) \oplus (0, j)$ representation, 
either at the elementary level  such as models for physics beyond the standard model with elementary high spin matter fields, 
or at the composite level in effective theories for hadronic interactions or in the description of composite objects in physics 
beyond the standard model.
   
The paper is organized as follows: in the next section we find all the representations having only two possible spin values, 
which are suitable to be used in the \poi\ projector formalism when we impose also to have at most a second order Lagrangian 
theory. In Section III we present the construction of the parity-based covariant basis for $(j, 0) \oplus (0, j)$ representations 
and their connection with the associated Jordan and Lie algebras. In Section IV we work out the consequences of this algebraic 
structure for the electromagnetic multipole moments of an elementary particle described by the \poi\ projector formalism 
and study the corresponding propagation of spin $j$ waves in an electromagetic background. Our conclusions are presented in  
Section V.

\section{General \poi\ projectors}

The \poi\ algebra is the semidirect product $\mathfrak{t}_4  \rtimes \mathfrak{so}(1,3)$, satisfying the Lie brackets
\begin{multline}
i[M\indices{_\mu_\nu} , P_{\rho}] = \eta\indices{_\mu_\rho} {P}_{\nu} - \eta\indices{_\nu_\rho} P_{\mu}, 
\qquad \qquad [{P}_\mu, {P}_\nu] = 0 \\ 
i[M\indices{_\mu_\nu}, M\indices{_\rho_\sigma}] = \eta\indices{_\mu_\rho} M\indices{_\nu_\sigma} 
- \eta\indices{_\mu_\sigma} M\indices{_\nu_\rho} - \eta\indices{_\nu_\rho} M\indices{_\mu_\sigma} 
+ \eta\indices{_\nu _\sigma} M\indices{_\mu_\rho},
\end{multline}

where indexes run as $\mu = 0,\ldots,3$. The $\so(1,3)$ subalgebra generated by the ${M}\indices{_\mu_\nu}$ is the 
Lorentz algebra. Is is customary to separate the Lorentz generators in \emph{rotations} $\vec{J}$ and 
\emph{boosts} $\vec{K}$ operators
\begin{equation}
M\indices{^0^i}=K_{i},\qquad M\indices{^i^j}=\epsilon\indices{_i_j_k}J_{k}.
\end{equation}

The \poi\ algebra has two algebraic invariants with the corresponding Casimir Operators given by
\begin{equation}
\cas{2}  = P^\mu P_\mu \qquad \cas{4} = W^\mu W_\mu.
\end{equation}
where $W_{\mu}$ stands for the Pauli-Lubanski four-vector 
\begin{equation}
W_\mu = \frac{1}{2} \varepsilon\indices{_\mu_\sigma_\tau_\rho}  M\indices{ ^\sigma ^\tau}P^\rho.
\end{equation}
The \poi\ one-particle states are then characterized by the quantum numbers coming from
\begin{equation}\label{eq:iso13qn}
\cas{2} \ket{\Psi} = m^2 \ket{\Psi} \qquad \cas{4} \ket{\Psi} = -m^2 j(j+1)\ket{\Psi}, 
\end{equation}
where $m$ denotes the mass and $j$ the spin. Now, the fields which enter our theories are linear combinations 
of the  creation and annihilation operators of these states, defined by the relation 
$\ket{m^2, j}=a^\dagger({m^2, j})\ket{0}$ and its adjoint. The transformation properties of the creation and 
annihilation operators under the \poi\ group are fixed by this relation, and in turn they fix (through 
\poi\ invariance of the scattering matrix and cluster decomposition) the general form for the fields as 
\cite{Weinberg:1995mt}
\begin{equation}\label{eq:fieldD}
\psi_l(x) = \int d \Gamma (\kappa e^{i p \cdot x} \omega_l(\Gamma) a^\dagger(\Gamma)+\lambda
e^{-i p \cdot x} \omega^{c}_l(\Gamma) a(\Gamma)),
\end{equation}
where $\Gamma$ is the set of labels $[m^2,j, p^\mu, \sigma]$, with $\sigma$ the quantum number of the little 
group (i.e., spin projection for massive fields or helicity for massless fields) and $\kappa$, $\lambda$ are 
constants. The coefficients $\omega_l$ and $\omega^{c}_l$ in this expression transform in some representation 
of the Lorentz group. In order to consider this construction complete, all that remains it to choose 
coefficients with the appropriate transformation rules and the constants $\kappa$, $\lambda$ with the appropriate
values to properly account for discrete symmetries.
 
Conventionally, we construct our field theories using only a handful of (in general, reducible) Lorentz 
representations: Dirac spinors, four vectors, spinor-vectors and higher order tensors for higher spin. 
We require, however, that these fields carry irreducible representations of the \poi\ group, that is, 
particles with definite mass and spin. Indeed, it is when this requisite is not properly satisfied that 
inconsistencies manifest, like superluminal propagation or the appearance of spurious degrees of freedom. 
The role of equations of motion is then to covariantly ensure that only the desired degrees of freedom are present. 

The homogeneous Lorentz algebra $\so(1,3)$ is locally isomorphic to the direct sum $\su(2)_{A}\oplus\su(2)_{B}$, 
spanned by the combinations of rotation $\vec{J}$ and boost $\vec{K}$ generators
\begin{equation}
\vec{A} =\frac{1}{2}( \vec{J} - i \vec{K}) \qquad \vec{B} = \frac{1}{2}(\vec{J} + i \vec{K}).
\end{equation}
These two sets commute
\begin{equation} 
[\vec{A}, \vec{B}] = 0.
\end{equation}
Therefore, we can label the irreducible Lorentz representations with two $\su(2)$ numbers  $(a,b)$. There are two 
Casimir operators for Lorentz group, $M\indices{_\mu _\nu} {M}\indices{^\mu ^\nu}$ and 
$M\indices{_\mu _\nu} \widetilde{M}\indices{^\mu ^\nu}$, where the dual tensor  $\widetilde{M}\indices{^\mu^\nu}$ is 
defined as 
\begin{equation}
\widetilde{M}\indices{^\mu^\nu}\equiv\frac{1}{2}\varepsilon\indices{^\mu^\nu^\alpha^\beta}M\indices{_\alpha_\beta}.
\end{equation}
These operators can be recast in terms of the $\su(2)$ Casimir operators $\vec{A}^2$ and $\vec{B}^2$ as
\begin{align}
M\indices{_\mu _\nu} M\indices{^\mu ^\nu} &= 4(\vec{A}^2 + \vec{B}^2 ) \\
M\indices{_\mu _\nu} \widetilde{M}\indices{^\mu ^\nu} &= -4i(\vec{A}^2 - \vec{B}^2 ).
\end{align}
The (linear, unitary) parity operator induce the following transformations of the Lorentz generators
\begin{align}
\Pi \vec{J} \Pi^{-1}& = \vec{J} \\
\Pi \vec{K} \Pi^{-1}& = - \vec{K},
\end{align}
so that $\vec{A}$ and $\vec{B}$ transform into each other
\begin{align}\label{eq:abparity}
\Pi \vec{A} \Pi^{-1}& = \vec{B} \\
\Pi \vec{B} \Pi^{-1}& = \vec{A}.
\end{align}
This immediately suggest a broad classification of the Lorentz representations (enlarged by parity) in two groups. 
First, the $(a,a)$ representations that transform into themselves under the action of parity, for which the second 
invariant is null, that is,   $M\indices{_\mu _\nu} \widetilde{M}\indices{^\mu ^\nu}=0$. These were the 
representations proposed by Fierz and Pauli to describe arbitrary integer spin $j=2a$ \cite{Fierz:1939ix}. Second, 
we have the  reducible $(a,b)\oplus(b,a)$ representations with $a\neq b$, for which $(a,b)$ and $(b,a)$ are exchanged 
by parity.  We call these \textit{chiral} representations, because for them we can define a chirality operator 
\begin{equation}
\chi =  \frac{i}{4a(a+1)-4b(b+1)}   M\indices{_\mu _\nu} \widetilde{M}\indices{^\mu ^\nu}. 
\label{chirality} 
\end{equation}
Since $\chi$ is proportional to a Casimir operator of the Lorentz group, we have the commutation rule
\begin{equation} \label{chim}
[\chi, M\indices{_\mu_\nu}]=0. 
\end{equation}
These chiral representations include the Joos-Weinberg representations \cite{Weinberg:1964cn} which correspond 
to $b=0$. 

The specific form of the projector selecting the degrees of freedom with quantum numbers $(m, j)$ depends on the 
spin content of the chosen representation. By working in Lorentz representations containing at most two spin sectors, 
we can build second-order projections in the momenta which can be implemented in Lagrangian form without the addition 
of constraints or auxiliary degrees of freedom \cite{Napsuciale:2006wr}.

In general, the representation $(a,b)$ contains states with all spins between $|a-b|$ and $a+b$. Those with at most 
two spin sectors which are also irreducible representations for parity can be enumerated as follows:
\begin{itemize}
\item The nonchiral representations $(0,0)$ and $\left(\frac{1}{2},\frac{1}{2} \right)$.
\item The single-spin chiral representations $(j,0)\oplus(0,j)$ with $j\geq\frac{1}{2}$.
\item The double-spin chiral representations $\left(j-\frac{1}{2},\frac{1}{2} \right) 
\oplus \left(\frac{1}{2},j-\frac{1}{2} \right)$ with $j> 1$. 
\end{itemize}	

For the single-spin $(j,0)\oplus(0,j)$ representation the \poi\ projector is 
\begin{equation} 
\mathcal{P}^{\{m,j\}} 
= \frac{P^{2}}{m^{2}} \left[  \frac{W^{2}}{-j(j+1)P^{2} }\right].  
\end{equation}
We remark that this projector fixes the appropriate mass and spin quantum numbers but in general other properties of 
the particles such as parity are not fixed.   

On the other hand, for the double-spin representations, we use instead the projector
\begin{equation} 
\mathcal{P}^{\{m,j\}} = \left[  \frac{W^{2}}{-2j m^{2}} - \frac{j(j-1)}{2j} \frac{P^{2}}{m^{2}}  \right]  
\end{equation}
which removes the unwanted spin $j-1$ components of the field to insure that only spin $j$ is propagated 
\cite{Napsuciale:2006wr}.

This projection produces second order equations of motion of the form 
\begin{equation} 
(T\indices{_\mu _\nu} P^{\mu}P^{\nu} - m^2)\Psi = 0 ,
\label{pp}
\end{equation}
for both fermions and bosons where the specific form of the tensor $T\indices{_\mu _\nu}$ depends on the chosen Lorentz 
representation. However, as discussed in the previous works 
\cite{Napsuciale:2006wr, Napsuciale:2007ry, DelgadoAcosta:2012yc}, only the symmetric part of the operator 
$T\indices{_\mu _\nu}$ is fixed by the projector. This requires  to construct the most general space-time antisymmetric 
operator which is clearly representation-dependent. Furthermore, a given Lorentz representation will house, 
in addition to these antisymmetric operators relevant for the description of gauge interactions, many other 
operators which can be relevant for non-gauge or self-interactions. In this work we aim to classify all of 
them for the restricted class of single-spin chiral $(j,0)\oplus(0,j)$ representations, in a construction based 
on the covariant properties of parity operator.

\section{Parity-based covariant basis for the \texorpdfstring{$(j,0) \oplus (0,j)$}{(j,0)+(0,j)}\ representation space.} 
The specific representation of operators depends on our choice for the basis, thus we start by fixing our 
conventions for the basis in $(j,0) \oplus(0,j)$ space. For the ``right'' representation $(j,0)$ we choose 
the angular momentum basis $\{|j,m\rangle_{R}  \}$.  Similarly, for the ``left'' representation $(0,j)$ we 
chose to work with the corresponding $\{|j,m\rangle_{L} \}$ basis. The Lorentz generators for the $(j,0)$ 
and $(0,j)$ representations are
\begin{align}
{M_{R}}\indices{^0^i}  &  =\left(  K_{R}\right)_{i},\qquad {M_{R}}\indices{^i^j}=\epsilon\indices{_i_j_k}\left( J_{R}\right)_{k},\\
{M_{L}}\indices{^0^i}  &  =\left(  K_{L}\right)_{i},\qquad {M_{L}}\indices{^i^j}=\epsilon\indices{_i_j_k} \left( J_{L}\right)_{k},
\end{align}
where $\vec{J}_R=\vec{J}_L=\boldsymbol\tau$ are the conventional $(2j+1)\times (2j+1)$ angular momentum matrices and 
$ \vec{K}_{R}=- \vec{K}_{L}=i \boldsymbol\tau$.  
With this choice the states $\{|j,m\rangle_{R} , |j,m\rangle_{L} \}$ form a basis for the direct sum 
$(j,0) \oplus(0,j)$ representation space which we will denote  as {\it chiral basis} in the following.  In this basis the 
Lorentz generators take the following form  
\begin{equation}
\vec{J}=\left( \begin{array}[c]{cc} \boldsymbol\tau & 0\\ 0 & \boldsymbol\tau \end{array} \right), \qquad \vec{K}=\left(
\begin{array}[c]{cc} i\boldsymbol\tau & 0\\ 0 & -i\boldsymbol\tau \end{array}
\right).
\end{equation}

The components of the Lorentz antisymmetric tensor for the $(j,0) \oplus(0,j)$ representation can be written in terms 
of these matrices as
\begin{equation}
M\indices{^i^j} = \epsilon\indices{_i_j_k} \begin{pmatrix} 
\tau_k &  0 \\ 
 0 & \tau_k
\end{pmatrix} \qquad
M\indices{^0^i} = i \begin{pmatrix} 
\tau_i & 0 \\ 
0 & -\tau_i
\end{pmatrix} .
\label{LG}
\end{equation}

From the general properties in \ref{eq:abparity}, we see that, in the chiral basis, the chirality operator in 
Eq. \ref{chirality} takes the diagonal form
\begin{equation}
\chi =  \begin{pmatrix} \mathbbm{1}_{2j+1}  & 0 \\ 0 & -\mathbbm{1}_{2j+1} \\\end{pmatrix}, 
\label{chi} 
\end{equation}
and the parity operator which 
swaps the chiral subspaces $\left( j, 0 \right)$ and $ \left(0,j\right)$ takes the form
\begin{equation}
\Pi = \begin{pmatrix}0 & \mathbbm{1}_{2j+1} \\ \mathbbm{1}_{2j+1} & 0 \\\end{pmatrix},
\label{Pi}
\end{equation} 
thus, parity and chirality anticommute for all $(j,0)\oplus(0,j)$ representations
\begin{equation}
 \{\Pi,\chi\} =0. 
\label{pichi}
\end{equation}

For these representations the Lorentz generators also satisfy
\begin{equation}
\vec{K}=i\chi\vec{J} \label{KJ}.
\end{equation}
which can be covariantly written as
\begin{equation}
\widetilde{M}\indices{^\mu^\nu} =-i\chi M\indices{^\mu^\nu}. \label{chiten}%
\end{equation}

Now, the parity operator fulfills the relations
\begin{align}
[\Pi,\vec{J}] &= 0 \\
[\Pi,\vec{K}] &= 2 \Pi \vec{K},
\end{align}
or in covariant form 
\begin{equation}
[M\indices{_\mu_\nu}, \Pi] = i \eta\indices{_0_\mu}( 2 i\Pi M\indices{_0_\nu}) - i \eta\indices{_0_\nu} (2 i\Pi M\indices{_0_\mu}).
\end{equation}

As these commutation rules make clear, parity, while rotating as a scalar, is not a Lorentz scalar under boosts.  
A straightforward calculation yields the following commutation rules for the object $V_k= 2i \Pi M\indices{_0_k}$
\begin{equation}\begin{split}\label{eq:pikcomm}
[M\indices{_i_j}, V_k] &=
-i \eta\indices{_i_k}  V_j + i \eta\indices{_j_k} V_i  \\
[M\indices{_0_i}, V_j] &=  -2\Pi \{M\indices{_0_i}, M\indices{_0_j}\}.
\end{split}\end{equation}
The composite object $V_k$ rotates as a vector but it does not behave as a vector under boosts. The transformation properties 
under boosts involve the anti-commutator $\{M\indices{_0_i}, M\indices{_0_j}\}$ (the Jordan algebra of Lorentz generators),  
which, in contrast to the Lie algebra, is not universal.  Writing Eqs. (\ref{eq:pikcomm}) in covariant notation we get
\begin{equation}\begin{split}\label{eq:picommcov}
[M\indices{_\mu_\nu}, V_\rho ]= {} & i(\eta\indices{_\nu_\rho} V_\mu -\eta\indices{_\mu_\rho} V_\nu) 
+ i \eta\indices{_0_\mu} \eta\indices{_0_\rho} V_\nu
- i \eta\indices{_0_\mu} \eta\indices{_0_\nu} V_\rho\\
& - i \eta\indices{_0_\nu} \eta\indices{_0_\rho} V_\mu 
+ i \eta\indices{_0_\mu} \eta\indices{_0_\nu}  V_\rho\\ 
& -2i \eta\indices{_0_\mu}\Pi\{M\indices{_0_\nu}, M\indices{_0_\rho}\}\\ 
& +2i \eta\indices{_0_\nu}\Pi \{M\indices{_0_\mu}, M\indices{_0_\rho} \} ).
\end{split}\end{equation}
The appearance of the quantity $\Pi \{M\indices{_0_\mu},M\indices{_0_\nu}\}$ on the right hand side suggests that in general 
the covariant properties of  of $\Pi$ will depend on the transformation properties of the objects 
\begin{align}
t_{\mu_1, \mu_{2}}&= \{ M\indices{_0_{\mu_1}}, M\indices{_0_{\mu_2}} \} ,...  \\
t_{\mu_{1} \ldots \mu_{2j}} &= \{  t_{\mu_1 \ldots \mu_{2j-1}}, M\indices{_0_{\mu_{2j}}} \}. 
\end{align}
By calculating, for a particular representation, the commutators with the Lorentz generators of the series 
$(\Pi,\Pi M\indices{_0_\mu}, \Pi t\indices{_{\mu_1}_{\mu_2}}, \ldots)$ we will eventually arrive at a set of 
objects transforming into themselves. It will be shown below that these operators form a symmetric tensor 
$S\indices{_{\mu_1}_{\ldots}_{\mu_{2j}}}$ whose time component $S\indices{_{0}_{\ldots}_{0}}$ is $\Pi$. 

In order to make our parity-based construction transparent, we start by studying the simplest case, the  
$\left(\frac{1}{2},0\right) \oplus \left(0,\frac{1}{2} \right)$ space which for obvious reasons we denote 
as Dirac representation in the following. Here and in the following, the inner product in the matrix space 
is taken to be 
\begin{equation}
A\cdot B = Tr[A B].
\end{equation} 

A basis for the operators acting on the $(j,0)\oplus (0,j)$ representation space can be obtained from the exterior 
product of states in the $\{|j,m\rangle_{R} ,|j,m\rangle_{L} \}$ basis. This set provides a basis for constructing 
the most general bilinear in the fields with definite Lorentz transformation properties. In particular for $j=1/2$, 
examining the Lorentz decomposition of the external product of states in this basis we get 
\begin{equation*}
\left[\left(\frac{1}{2},0\right) \oplus \left(0,\frac{1}{2}\right) \right]^2 = 
(0,0)_2 \oplus(1,0) \oplus(0,1) \oplus \left(\frac{1}{2}, \frac{1}{2}\right)_2. 
\end{equation*} 
In this equation, the left side stands for the exterior product of the basis states, in the right side we have the 
Lorentz decomposition of this product, with subscripts denoting multiplicity. It corresponds to a pair of scalars, 
an antisymmetric tensor, and a pair of four-vectors. 

The first operator transforming in the $(0,0)$ representation is the unit operator $\mathbbm{1}$. Given that 
the chirality operator commutes with the Lorentz generators (\ref{chim}), we can use this operator as the second 
operator transforming in the $(0,0)$ representation.
Operators transforming in the $(1,0)\oplus(0,1)$ are clearly the Lorentz generators $M_{\mu\nu}$. The remaining 
operators can be built analyzing the covariant properties of parity.

Recalling the transformation rules \ref{eq:picommcov}, we need to construct the $t\indices{_{i}_{j}}$ operators. 
For the Dirac representation, $\{M\indices{_0_i}, M\indices{_0_j}\} = \frac{1}{2} \delta\indices{_i_j}$. Defining
the object
\begin{equation}
S_\mu = \eta\indices{_0_\mu}\Pi - 2i\Pi M\indices{_0_\mu},
\label{gammamu}
\end{equation} 
we can see that it transforms as a four-vector
\begin{equation}\label{eq:gammalorentz}
[M\indices{_\rho_\sigma}, S_\mu ] = i \eta\indices{_\mu_\rho} S_\sigma -  i \eta\indices{_\mu_\sigma} S_\rho, 
\end{equation} 
thus we conclude that $\Pi$ transforms as the zeroth-component of the four-vector $S^{\mu}$. 
   
An important property induced by Eqs. (\ref{chim},\ref{pichi}) and the specific form of $S^{\mu}$ in Eq. (\ref{gammamu}) is 
\begin{equation}
\{ \chi, S^{\mu}\}=0,
\label{chigamma}
\end{equation}
which implies the orthogonality of $\chi$ and $S_\mu$. A simple combination of Eqs. (\ref{chim}, \ref{eq:gammalorentz}) 
shows that $\chi S_\mu$ transforms also as a four-vector. Finally, a direct calculation yields that these are independent operators. 

In summary, for the $(1/2,0)\oplus (0,1/2)$ representation our parity-based construction yields the following covariant basis
\begin{equation}\label{covbasis12}
\left\{ \mathbbm{1},\chi, S_\mu, \chi S_\mu, M\indices{_\mu_\nu} \right\}.
\end{equation} 

It is well-known that the conventional sixteen matrices
\begin{equation}\label{eq:diraccontent}
\left\{ \mathbbm{1},\gamma_5, \gamma_\mu, \gamma_5\gamma_\mu, \sigma\indices{_\mu_\nu} \right\},
\end{equation} 
form a basis for this operator space, with the Lorentz generators for this representation space given by 
$M\indices{_\mu_\nu}  = \frac{1}{2} \sigma\indices{_\mu_\nu} $. A direct comparison shows that our procedure reproduces the 
conventional covariant basis with the $\gamma^{\mu}$ matrices in the Weyl representation, except for an irrelevant $1/2$ 
normalization factor in $M_{\mu\nu}$. It  is also clear that the specific form of the operators in Eq. (\ref{covbasis12}) 
depend on the choice of the basis we use for the states in $(1/2,0)\oplus (0,1/2)$.
 
In terms of the covariant operator $S^\mu$ in Eq. (\ref{gammamu}), the rest-frame parity projection equation
\begin{equation}
\frac{1}{2}(1\pm \Pi ) \psi(0) =\psi(0)
\end{equation}
transforms,  for an arbitrary frame,  into the familiar
\begin{equation}
(S_\mu P^\mu \mp m) \psi(p) = 0.
\end{equation}
From this perspective, the Dirac equation is simply the covariant projection over parity-invariant subspaces in the 
$(1/2,0)\oplus (0,1/2)$ representation space and the Dirac algebra satisfied  by the $S^{\mu}$ matrices
\begin{equation}
\{ S^{\mu},S^{\nu} \}=2\eta\indices{^\mu^\nu},
\label{Jordan12}
\end{equation}
is just a manifestation of the covariant properties of the parity operator.

Using the projection over states with well-defined spin and mass produces a condition of the general form in Eq. (\ref{pp}). 
For the $(1/2,0)\oplus (0,1/2)$ representation, the symmetric part of the $T\indices{_\mu_\nu}$ space-time tensor is fixed 
by the projector to be $\eta\indices{_\mu_\nu}$, but  the antisymmetric part is not, thus its general form must be 
constructed in terms of the covariant basis. For this representation the only possibility is a term $g M\indices{_\mu_\nu}$, 
with $g$ arbitrary, which upon gauging produces an interaction 
\begin{equation}
 g M\indices{_\mu_\nu} F\indices{^\mu^\nu}.
\end{equation}
Such a formalism has been studied at one-loop for the abelian and nonabelian 
cases~\cite{AngelesMartinez:2011nt,VaqueraAraujo:2012qa, Vaquera-Araujo:2013bwa}.

\subsection{Lorentz structure of the operators acting on \texorpdfstring{$(1,0) \oplus (0,1)$}{(1,0)+(0,1)}. }
 
As a second example, let us proceed in full detail with the $(1,0)\oplus(0,1)$ construction, which then we can 
generalize to arbitrary $j$. The conventional angular momentum generators entering Eq.(\ref{LG}) for this 
representation are 
\begin{equation}\begin{split}
&\tau_1 = \frac{1}{2}\begin{pmatrix} 
0 & \sqrt{2} & 0 \\ \sqrt{2} & 0 & \sqrt{2} \\ 0 & \sqrt{2} & 0 
\end{pmatrix} \\
&\tau_2 = \frac{i}{2}\begin{pmatrix} 
0 & -\sqrt{2} & 0 \\ \sqrt{2} & 0 & -\sqrt{2} \\ 0 & \sqrt{2} & 0 
\end{pmatrix} \\
&\tau_3 = \frac{1}{2}\begin{pmatrix}
 2 & 0 & 0 \\ 0 & 0 & 0 \\  0 & 0 & -2 \end{pmatrix}.
\end{split}\end{equation}

A basis for the operators acting on the $(1,0)\oplus(0,1)$ space can be obtained via the external products of the 
states in the $\{|j,m\rangle_{R} ,|j,m\rangle_{L} \}$ basis which has the following Lorentz decomposition
\begin{equation}\label{eq:j1decom}
\begin{split}
\left[(1,0) \oplus (0,1)\right]^2 = {} &  (0,0)_2 \oplus (1,1)_2  \oplus (1,0) \oplus (0,1) \\ & \oplus(2,0)\oplus(0,2)
\end{split}
\end{equation}
We may identify the two operators transforming in the $(0,0)$ representation as the unit and chirality operators, and 
the Lorentz generators with the  operators transforming in the $ (1,0) \oplus (0,1)$. However, it is not obvious how 
can we construct  the operators transforming in the two $(1,1)$ and the $(2,0)\oplus(0,2)$ Lorentz representations. 
In order to construct these operators and aiming to envision the general case, we briefly review the theory of 
$\so(1,3)$ Young projectors. 

An arbitrary traceless Lorentz tensor of rank $r$ can be  decomposed in an orthogonal basis given by all the 
completely traceless tensors enumerated by all possible Young tableaux of $r$ boxes. These Young tableaux  
index the representations of the symmetric group $S_N$. 
We identify them with the permutation properties of the Lorentz indexes of our tensor. For example, a symmetrical 
tensor $S\indices{_\mu_\nu}$ of rank $2$ corresponds to a row Young tableau 
$\begin{ytableau} \mu & \nu \\ \end{ytableau}$, while an antisymmetrical tensor corresponds 
to the column Young tableau\    $\begin{ytableau} \mu \\ \nu \\ \end{ytableau}$\ . 

A Young projector associated with some Young tableau, is an operator which projects a general tensor into the 
subspace with the symmetries of the tableau. Since these projectors are built with the metric tensor 
$\eta\indices{_\mu_\nu}$, which is an invariant tensor of the Lorentz algebra, the Young projection is also 
invariant. Therefore, the subspaces transform separately. This fact is at the root of our decomposition.

We are interested in the chiral representations, which produce either totally symmetrical or self-dual/anti-self 
dual tensors. We only need the characterization of the Young diagrams with one and two rows \cite{Qi-Ma:2007}. 
The dimension of a completely traceless tensor (i.e. a tensor for which every contraction vanishes) in $\so(1,3)$ 
is given by the following combinatorial formulas 
\begin{align}
\begin{split}
d_{[n]} = (n+1)^2 \quad 
&\begin{ytableau} \mu_1 &\mu_2 & \mu_3 & \none[\ \ldots\ ] & \mu_n \\ \end{ytableau}
\end{split} \\
\begin{split}
d_{[n,m]} = 
2n^2+4n-2m^2+2 \quad
&\begin{ytableau} \mu_1 & \mu_2 & \mu_3 & \none[\ \ldots\ ] & \mu_n \\ \nu_1 & \nu_2 & \none[\ldots] & \nu_m \\ \end{ytableau}
\end{split}.
\end{align}
These diagrams describe either a symmetrical tensor of rank $n$, or a mixed-symmetry tensor of rank $n+m$. The 
factor $2$ comes about because we are considering both the self-dual and anti-self dual parts. (For an in-depth 
discussion of this technical point, see chapter 9 of \cite{Qi-Ma:2007}).

The Young projector corresponding to a given Young pattern is constructed as the product of the appropriate 
symmetrizers and antisymmetrizers. For row or column Young tableau, we have the pure symmetrizers and 
antisymmetrizers, while for mixed-symmetry tableaux we chose to first antisymmetrize index subsets, and then 
to symmetrize as appropriate. This choice is not unique, but is convenient for calculations 
\cite{Elvang:2003ue, Cvitanovic:2008zz}. For example, to obtain the Young projector corresponding to the diagram: 
\begin{equation}
\begin{ytableau} \mu&\alpha \\ \nu&\beta \\ \end{ytableau} 
\end{equation}
we first antisymmetrize the column pairs, and then symmetrize the row pairs:
\begin{equation}
\mathbb{P}\indices{^\mu ^\nu ^\alpha ^\beta _\xi_\zeta_\lambda_\kappa} \propto 
\mathcal{S}\indices{^\mu ^\alpha_\rho_\sigma} \mathcal{S}\indices{^\nu ^\beta_\tau_\delta} 
\mathcal{A}\indices{^\rho^\tau_\xi _\zeta} \mathcal{A}\indices{^\sigma^\delta_\lambda _\kappa},
\end{equation}
with a suitable normalization.

Coming back to the $(1,0)\oplus(0,1)$ representation, we again have two scalars, which are chosen as the unit 
operator and the chirality operator $\chi$. We expect that the rest of the covariant basis will be decomposable as:
\begin{equation}
2\ \ydiagram{2}  \oplus \ydiagram{1,1} \oplus \ydiagram{2,2}. 
\end{equation}
This result tells us that the basis we seek is provided by the following operators
\begin{equation}
\left\{ 1,\chi, S\indices{_\mu_\nu},  \mathcal{S}\indices{_\mu_\nu}, M\indices{_\mu_\nu}, C\indices{_\mu_\nu _\alpha_\beta} \right\},
\label{s1basis}
\end{equation} 
where $S\indices{_\mu_\nu}$ is a symmetric traceless ($S\indices{^\mu_\mu}=0$) tensor operator thus having $9$ 
independent components which coincide with the degrees of freedom of the $(1,1)$ Lorentz representation. Similar 
results hold for the $ \mathcal{S}\indices{_\mu_\nu}$ tensor operator which also has 9 independent components. 

As for the $ C\indices{_\mu_\nu _\alpha_\beta}$ tensor operator, it  has the following symmetries %
\begin{equation}
C_{\mu\nu\alpha\beta}=-C_{\nu\mu\alpha\beta}=-C_{\mu\nu\beta\alpha},\quad
C_{\mu\nu\alpha\beta}=C_{\alpha\beta\mu\nu}, \label{Weylsym}%
\end{equation}
it is traceless, i.e., the contraction of any pair of indices vanishes, and it satisfies the algebraic Bianchi identity%
\begin{equation}
C_{\mu\nu\alpha\beta}+C_{\mu\alpha\beta\nu}+C_{\mu\beta\nu\alpha}=0.
\label{Bianchi}%
\end{equation}
A fourth rank tensor has $256$ components but it is easy to convince ourselves that these symmetries restrict 
this tensor to have only $10$ independent components, which are precisely the number of degrees of freedom in 
the $(2,0) \oplus (0,2)$ Lorentz representation. The calculation is simple, and familiar from the Riemann tensor: 
our tensor it is akin to a rank $2$ symmetrical tensor in six dimensions, which has $21$ independent components, 
but the traceless condition removes ten, and the Bianchi identity removes one, leaving the aforestated ten independent 
components.

Aiming to construct explicitly the tensors in the operator basis in Eq. (\ref{s1basis}), we calculate the Lorentz 
commutators of the series $\Pi, \Pi M\indices{_0 _i},  \Pi t\indices{_i_j}, \ldots$
\begin{equation}\begin{split}
[M\indices{_\mu_\nu}, \Pi] = {} & -2 \eta\indices{_0_\mu}  \Pi M\indices{_0_\nu} 
- 2 \eta\indices{_0_\nu} \Pi M\indices{_0_\mu} \\
[M\indices{_\mu_\nu},  \Pi M\indices{_0_i}] =& -i \eta\indices{_\mu_i}  \Pi M\indices{_0_\nu} 
+  i \eta\indices{_\nu_i} \Pi M\indices{_0_\mu} \\ 
& +  \eta\indices{_0_\mu}  \Pi t\indices{_\nu_i} - \eta\indices{_0_\nu}  \Pi t\indices{_\mu_i} \\
[M\indices{_\mu_\nu},  \Pi t\indices{_i_j} ] = {} & - \eta\indices{_0_\mu} \Pi t\indices{_\nu_i_j} 
+ \eta\indices{_0_\nu} \Pi t\indices{_\mu_i_j} \\ 
&-i \eta\indices{_\mu _i} \Pi t\indices{_\nu_j} +i \eta\indices{_\nu _i} \Pi t\indices{_\mu_j}  \\  
&  -i \eta\indices{_\mu _j} \Pi t\indices{_\nu_i} +i \eta\indices{_\nu _j} \Pi t\indices{_\mu_i} .
\end{split}\end{equation}

For this representation the following relation holds
\begin{equation}
t\indices{_\rho_\mu_\nu}= \eta\indices{_\mu_\rho } M\indices{_0_\nu} +  \eta\indices{_\nu_\rho } M\indices{_0_\mu} 
+ 2 \eta\indices{_\mu_\nu } M\indices{_0_\rho} . 
\end{equation}

The conclusion is that the operators $( \Pi, \Pi M\indices{_0_\mu}, \Pi \{M\indices{_0_\mu}, M\indices{_0_\nu}\})$ 
transform into themselves under the Lorentz group, and therefore the symmetric traceless tensor $S\indices{^\mu^\nu}$ 
must be constructed as a linear combinations of these operators. It is easy to check that the appropriate combination is
\begin{equation}
S\indices{_\mu_\nu}=\Pi \eta\indices{_\mu_\nu}-i \Pi(\eta\indices{_0_\mu} M\indices{_0_\nu} 
+ \eta\indices{_0_\nu}M\indices{_0_\mu}) - \Pi \{M\indices{_0_\mu}, M\indices{_0_\nu}\}.
\end{equation} 
These operators are also traceless in the ``spinor'' space, and consequently orthogonal to the unit operator. 
Using Eqs. (\ref{chim},\ref{pichi}) it  is easy to show that 
\begin{equation}
\{ \chi, S\indices{^\mu^\nu} \}=0,
\label{chis}
\end{equation}
which yields
\begin{equation}
Tr(\chi S\indices{^\mu^\nu})=0,
\end{equation}
thus $\chi$ and $S\indices{^\mu^\nu}$  are also orthogonal operators. 

A straightforward calculation yields the following commutation relations
\begin{equation}\begin{split} \label{crms}
i \left[ M\indices{^\mu^\nu}, S\indices{^\alpha^\beta} \right] = {} &  \eta\indices{^\mu^\alpha} S\indices{^\nu^\beta} 
- \eta\indices{^\nu^\alpha} S\indices{^\mu^\beta} + \eta\indices{^\mu^\beta} S\indices{^\nu^\alpha} 
- \eta\indices{^\nu^\beta} S\indices{^\mu^\alpha}  \\ 
\left\{ M\indices{^\mu^\nu}, S\indices{^\alpha^\beta} \right\}  = {} &  \varepsilon\indices{^\mu^\nu ^\sigma^\beta} \chi S\indices{^\alpha_\sigma} + \varepsilon\indices{^\mu^\nu^\sigma^\alpha} \chi S\indices{^\beta_\sigma} \\ 
i\lbrack S\indices{^\mu^\nu},S\indices{^\alpha^\beta}] = {} &  \eta\indices{^\mu^\alpha} M\indices{^\nu^\beta} 
+ \eta\indices{^\nu^\alpha} M\indices{^\mu^\beta}  \\ &  + \eta\indices{^\nu^\beta} M\indices{^\mu^\alpha} 
+  \eta\indices{^\mu^\beta} M\indices{^\nu^\alpha}. 
\end{split}\end{equation}
Using these relations and Eq. (\ref{chis}) it is possible to show that 
$\chi S\indices{_\mu_\nu}$ is also 
an independent set of nine orthogonal operators, thus  the second traceless symmetric tensor is given by 
$\mathcal{S}\indices{_\mu_\nu}=\chi S\indices{_\mu_\nu}$. 

Finally, the fourth-order Weyl-like tensor $C\indices{_\mu_\nu _\alpha_\beta}$ can be built from the product 
$M\indices{_\mu_\nu}M\indices{_\alpha_\beta}$ by applying the projection operator 
\begin{equation}
\begin{ytableau} \mu&\alpha \\ \nu&\beta \\ \end{ytableau}  \propto \mathcal{S}\indices{_\mu _\alpha} 
\mathcal{S}\indices{_\nu _\beta} \mathcal{A}\indices{_\mu _\nu} \mathcal{A}\indices{_\alpha _\beta}
\end{equation}
which gives
\begin{equation}
T\indices{_\mu_\nu_\alpha_\beta}= 4\{M\indices{_\mu_\nu},M\indices{_\alpha_\beta} \}
+ 2\{M\indices{_\mu_\alpha},M\indices{_\nu_\beta} \} - 2\{M\indices{_\mu_\beta},M\indices{_\nu_\alpha} \}.
\end{equation}
Removing the Young-projected trace of this tensor we get
\begin{equation}\begin{split}
C\indices{_\mu_\nu_\alpha_\beta}  = {} & T\indices{_\mu_\nu_\alpha_\beta} + \frac{1}{2}(\eta\indices{_\mu_{\left[\alpha\right.}} T\indices{_{\left.\beta\right]}_{\nu}} -\eta\indices{_\nu_{\left[\alpha\right.}} T\indices{_{\left.\beta\right]}_{\mu}} ) \\ 
& - \frac{1}{6} \eta\indices{_\mu_{\left[\alpha\right.}}  \eta\indices{_{\left.\beta\right]}_{\nu}} T\indices{_\rho^\rho} \\
 = {} & T\indices{_\mu_\nu_\alpha_\beta} - 8 (\eta\indices{_\mu_\alpha}\eta\indices{_\nu_\beta}
  -\eta\indices{_\nu_\alpha}\eta\indices{_\mu_\beta}) 
\end{split}\end{equation}
where
\begin{equation}
T\indices{_\mu_\nu} = T\indices{_\mu_\beta_\nu^\beta}.
\end{equation}
Finally, the following tensor
\begin{equation}\begin{split}
C\indices{_\mu_\nu_\alpha_\beta} = {} & 4\{M\indices{_\mu_\nu}, M\indices{_\alpha_\beta} \} 
+ 2\{M\indices{_\mu_\alpha}, M\indices{_\nu_\beta} \} \\ & - 2\{M\indices{_\mu_\beta}, M\indices{_\nu_\alpha} \}  
- 8 (\eta\indices{_\mu_\alpha} \eta\indices{_\nu_\beta} - \eta\indices{_\nu_\alpha} \eta\indices{_\mu_\beta})
\end{split} \end{equation}
obeys the symmetries in Eqs. (\ref{Weylsym}) and the Bianchi identity in Eq. (\ref{Bianchi}). These relations 
together with the vanishing of all contractions
\begin{equation}
C\indices{_\mu^\beta_\alpha_\beta} = 0,
\label{Ctensor}
\end{equation} 
leave only ten independent components. A direct calculation shows that this set is orthogonal to the 
previously constructed operators. 

Summarizing our construction for this representation, the parity-based covariant basis for 
$(1,0)\oplus (0,1)$ is given by the set
\begin{equation}
\left\{ 1,\chi, S\indices{_\mu_\nu},  \chi S\indices{_\mu_\nu}, M\indices{_\mu_\nu}, C\indices{_\mu_\nu _\alpha_\beta} \right\}. 
\label{bvbasis}
\end{equation} 

Concerning the \poi\ projector formalism, the most general antisymmetric tensor $T\indices{_\mu_\nu}$ for 
this representation can be expanded in terms  of the operators in the basis in Eq. (\ref{bvbasis}). Beyond 
$M\indices{_\mu_\nu}$ the only possibilities are the contractions 
$C\indices{_\mu_\nu _\alpha_\beta} M\indices{^\alpha^\beta}$ and 
$C\indices{_\mu_\nu _\alpha_\beta} \eta\indices{^\alpha^\beta}$. The latter contraction 
vanishes due to the properties of the $C$ tensor and the former is not independent and in turn can be expanded 
in terms of the covariant basis. Since the only antisymmetric tensor with the appropriate symmetries is the Lorentz 
generator tensor, this contraction must be proportional to $M\indices{_\mu_\nu}$. In summary, the most general 
antisymmetric tensor for $(1,0)\oplus (0,1)$ representation space is of the form $g M\indices{_\mu_\nu}$.

It is clear that high spin brings into the construction commutators and anti-commutators of the generators and it 
is important to realize the algebraic structure of these operators. We start with the Dirac space where the following 
commutation  relations hold 
\begin{equation}\begin{aligned}[]
[S_\mu,S_\nu] & = 4 i M_{\nu \mu} & 
[\chi S_\mu,\chi S_\nu] &=  \  4 i M_{\mu \nu} \\
[\chi S_\mu, S_\nu] &= \  2 \eta_{\mu \nu} \chi &
[\chi,S_\mu] &=   2 \chi S_{\mu} \\
[\chi,\chi S_\mu] &=   2 S_{\mu}. 
\end{aligned}\end{equation}
For anticommutators, we get the following results 
\begin{equation}\begin{aligned}
\{S_\mu,S_\nu\} &= 2 \eta_{\mu \nu} \mathbbm{1} & 
\{\chi,S_\mu\} &=  0  \\
\{\chi S_\mu,\chi S_\nu\} &=  -2 \eta_{\mu \nu} \mathbbm{1} &
\{\chi,\chi S_\mu\}  &=  0 & \\
\{\chi S_\mu, S_\nu\} &=  4 \widetilde{M}_{\mu \nu} &
\{\chi,\chi \} &=  2 \mathbbm{1}\\
\{S_\rho, M_{\mu \nu}\} &= -\varepsilon_{\rho \mu \nu \alpha} \chi S^\alpha & 
\{\chi, M_{\mu \nu} \} &=  2 i \widetilde{M}_{\mu \nu}  \\
\{\chi S_\rho,M_{\mu \nu} \} &= - \varepsilon_{\rho \mu \nu \alpha} S^\alpha.
\end{aligned}\end{equation}
These relations are schematically summarized in tables \ref{tb:diraclie} and \ref{tb:diracjordan}.
\begin{table}[tbp] \centering 
\caption{Algebraic Lie structure of the Dirac basis.}\label{tb:diraclie}
\begin{ruledtabular} 
\begin{tabular}{ c  c c c c c }
Lie $[\cdot,\cdot]$ &   $\mathbbm{1}$ & $\chi$ &  $S_\mu$ & $\chi S_\mu$ & $M\indices{_\mu_\nu}$ \\ 
\colrule
$\mathbbm{1}$ & $0$ & $0$ & $0$ & $0$ & $0$\\ 
$\chi$ & $0$ & $0$ &  $\chi S_\mu$ & $S_\mu$ & $0$\\ 
$S_\mu$ & $0$ & $\chi S_\mu$ &  $M\indices{_\mu_\nu}$ & $\chi$ & $S_\mu$\\ 
$\chi S_\mu$ & $0$ & $S_\mu$ & $\chi$ &  $M\indices{_\mu_\nu}$ &  $\chi S_\mu$\\ 
$M\indices{_\mu_\nu}$ & $0$ & $0$ &  $S_\mu$ & $\chi S_\mu$ & $M\indices{_\mu_\nu}$\\ 
\end{tabular} \end{ruledtabular}
\end{table} 

\begin{table}[tbp] \caption{Algebraic Jordan structure of the Dirac basis.}\label{tb:diracjordan}
\centering \begin{ruledtabular}
\begin{tabular}{ c c c c c c }
Jordan $\{\cdot , \cdot\}$  &   $\mathbbm{1}$ & $\chi$ &  $S_\mu$ & $\chi S_\mu$ & $M\indices{_\mu_\nu}$ \\ 
\colrule
$\mathbbm{1}$ & $\mathbbm{1}$ & $\chi$ &  $S_\mu$ & 
$\chi S_\mu$ & $M\indices{_\mu_\nu}$ \\ 
$\chi$ & $\chi$ & $\mathbbm{1}$ &  $0$ & $0$ & $M\indices{_\mu_\nu} $\\ 
$S_\mu$ &  $S_\mu$ & $0$ &  $\mathbbm{1}$ & $M\indices{_\mu_\nu}$ & $\chi S_\mu$\\ 
$\chi S_\mu$ & $\chi S_\mu$ & $0$   &  $M\indices{_\mu_\nu}$ & $\mathbbm{1}$ &  $S_\mu$\\ 
$M\indices{_\mu_\nu}$ & $M\indices{_\mu_\nu}$ & $M\indices{_\mu_\nu}$ &  $\chi S_\mu$ & $S_\mu$ & $\mathbbm{1}, \chi$\\ 
\end{tabular} \end{ruledtabular}        
\end{table}

Let us consider now the covariant basis we have constructed for the $(1,0)\oplus(0,1)$ representation. Besides the 
commutation rules in Eqs. (\ref{crms}), a straightforward calculation yields the following Lie brackets
\begin{equation}\begin{split}
[\chi S\indices{_\mu_\nu},\chi S\indices{_\mu_\nu}] = {} &  i \eta\indices{_\mu_\rho} M\indices{_\nu_\sigma} 
+ i\eta\indices{_\nu_\rho} M\indices{_\mu_\sigma} \\&
+ i\eta\indices{_\mu_\sigma} M\indices{_\nu_\rho} 
+ i\eta\indices{_\nu_\sigma} M\indices{_\mu_\rho}  \\ 
[\chi S\indices{_\mu_\nu}, S\indices{_\rho_\sigma}] = {} &  
\frac{4}{3}\left(\eta\indices{_\mu_\rho} \eta\indices{_\nu_\sigma} + \eta\indices{_\mu_\sigma} \eta\indices{_\nu_\rho} 
- \frac{1}{2} \eta\indices{_\mu_\nu} \eta\indices{_\rho_\sigma}\right) \\ & - \frac{i}{6} \left(\widetilde{C}\indices{_\mu_\rho_\nu_\sigma} 
+ \widetilde{C}\indices{_\mu_\sigma_\nu_\rho}  \right) \\
[\chi, S\indices{_\mu_\nu}] = {} &  2 \chi S\indices{_\mu_\nu}  \\
[\chi, \chi S\indices{_\mu_\nu}] = {} &  2 S\indices{_\mu_\nu}, 
\end{split}\end{equation}
and the anticommutators
\begin{equation}\begin{split}
\{ S\indices{_\mu_\nu},S\indices{_\rho_\sigma} \} = {} & \frac{4}{3}\left(\eta\indices{_\mu_\rho} \eta\indices{_\nu_\sigma} 
+ \eta\indices{_\mu_\sigma} \eta\indices{_\nu_\rho} -\frac{1}{2} \eta\indices{_\mu_\nu} \eta\indices{_\rho_\sigma}\right) \\ 
& - \frac{1}{6} \left({C}\indices{_\mu_\rho_\nu_\sigma} + {C}\indices{_\mu_\sigma_\nu_\rho}  \right)\\
\{\chi S\indices{_\mu_\nu},\chi S\indices{_\mu_\nu}\} = {} &  
-\frac{4}{3}\left(\eta\indices{_\mu_\rho} \eta\indices{_\nu_\sigma} + \eta\indices{_\mu_\sigma} \eta\indices{_\nu_\rho} 
-\frac{1}{2} \eta\indices{_\mu_\nu} \eta\indices{_\rho_\sigma}\right) \\
&+ \frac{1}{6} \left({C}\indices{_\mu_\rho_\nu_\sigma} + {C}\indices{_\mu_\sigma_\nu_\rho}  \right)\\ 
\{\chi S\indices{_\mu_\nu}, S\indices{_\rho_\sigma}\} = {} &  
\frac{1}{2}\left( \eta\indices{_\mu _\rho} \widetilde{M}\indices{_\nu _\sigma} 
+ \eta\indices{_\nu _\sigma} \widetilde{M}\indices{_\mu _\rho}\right) \\ 
& + \frac{1}{2} \left( \eta\indices{_\mu _\sigma} \widetilde{M}\indices{_\nu _\rho} 
+ \eta\indices{_\nu _\rho} \widetilde{M}\indices{_\mu _\sigma} \right) \\
\{M\indices{_\mu_\nu}, M\indices{_\rho_\sigma}\} = {} & \frac{4}{3} \left( \eta\indices{_\mu_\rho} \eta\indices{_\nu_\sigma} 
- \eta\indices{_\mu_\sigma} \eta\indices{_\nu_\rho}  \right) \\ & - \frac{8}{6} i \varepsilon\indices{_\mu_\nu_\rho_\sigma} \chi 
+ \frac{1}{6} C\indices{_\mu_\nu_\rho_\sigma}\\
\{\chi, S\indices{_\mu_\nu}\} = {} & 0  \\
\{\chi, \chi S\indices{_\mu_\nu}\} = {} & 0.
\end{split}\end{equation}
Here $\widetilde{C}\indices{_\mu_\rho_\nu_\sigma} = \frac{1}{2} \epsilon \indices{_\mu_\rho ^\alpha^\beta} {C}\indices{_\alpha_\beta_\nu_\sigma} 
= - i \chi {C}\indices{_\mu_\rho_\nu_\sigma}$. The similarities with the Dirac case can best be seen in the tables \ref{tb:j1lie} 
and \ref{tb:j1jordan}.
\begin{table*}[tbp] 
\caption{Algebraic Lie structure of the $(1,0)\oplus(0,1)$ basis.} \label{tb:j1lie}
\centering  \begin{ruledtabular}
\begin{tabular}{ c c c c c c c }
Lie $[\cdot,\cdot]$ &   $\mathbbm{1}$ & $\chi$ &  $S\indices{_\mu_\nu}$ & $\chi S\indices{_\mu_\nu}$ 
& $M\indices{_\mu_\nu}$ & $C\indices{_\mu_\nu_\rho_\sigma}$  \\ 
\colrule
$\mathbbm{1}$ & $0$ & $0$ & $0$ & $0$ & $0$ & $0$\\ 
$\chi$ & $0$ & $0$ &  $\chi S\indices{_\mu_\nu}$ & $S\indices{_\mu_\nu}$ & $0$ & $0$\\ 
$S\indices{_\mu_\nu}$ & $0$ & $\chi S\indices{_\mu_\nu}$ &  $M\indices{_\mu_\nu}$ & $\chi, C\indices{_\mu_\nu_\rho_\sigma}$ 
& $S\indices{_\mu_\nu}$ & $\chi S\indices{_\mu_\nu}$   \\ 
$\chi S\indices{_\mu_\nu}$ & $0$ & $S\indices{_\mu_\nu}$ & $\chi, C\indices{_\mu_\nu_\rho_\sigma}$ &  $M\indices{_\mu_\nu}$ 
&  $\chi S\indices{_\mu_\nu}$ & $S\indices{_\mu_\nu}$ \\ 
$M\indices{_\mu_\nu}$ & $0$ & $0$ &  $S\indices{_\mu_\nu}$ & $\chi S\indices{_\mu_\nu}$ & $M\indices{_\mu_\nu}$ 
& $C\indices{_\mu_\nu_\rho_\sigma}$\\ 
$C\indices{_\mu_\nu_\rho_\sigma}$ & $0$ & $0$ &  $\chi S\indices{_\mu_\nu}$ &  $S\indices{_\mu_\nu}$ 
& $C\indices{_\mu_\nu_\rho_\sigma}$ &  $M\indices{_\mu_\nu}$\\ 
\end{tabular}  \end{ruledtabular}
\end{table*}
\begin{table*}[tbp] \centering 
\caption{Algebraic Jordan structure of the $(1,0)\oplus(0,1)$ basis.}\label{tb:j1jordan}
\begin{ruledtabular}
\begin{tabular}{ c c c c c c c }
Jordan $\{\cdot,\cdot\}$  &   $\mathbbm{1}$ & $\chi$ &  $S\indices{_\mu_\nu}$ & $\chi S\indices{_\mu_\nu}$ 
& $M\indices{_\mu_\nu}$ & $C\indices{_\mu_\nu_\rho_\sigma}$ \\ 
\colrule
$\mathbbm{1}$ & $\mathbbm{1}$ & $\chi$ &  $S\indices{_\mu_\nu}$ & 
$\chi S\indices{_\mu_\nu}$ & $M\indices{_\mu_\nu}$ & $C\indices{_\mu_\nu_\rho_\sigma}$ \\ 
$\chi$ & $\chi$ & $\mathbbm{1}$ &  $0$ & $0$ & $M\indices{_\mu_\nu}$ & $C\indices{_\mu_\nu_\rho_\sigma}$ \\ 
$S\indices{_\mu_\nu}$ &  $S\indices{_\mu_\nu}$ & $0$ &  $\mathbbm{1}, C\indices{_\mu_\nu_\rho_\sigma}$ 
& $M\indices{_\mu_\nu}$ & $\chi S\indices{_\mu_\nu}$ & $S\indices{_\mu_\nu}$\\ 
$\chi S\indices{_\mu_\nu}$ & $\chi S\indices{_\mu_\nu}$ & $0$   &  $M\indices{_\mu_\nu}$ & $\mathbbm{1}, C\indices{_\mu_\nu_\rho_\sigma}$ 
&  $S\indices{_\mu_\nu}$ & $\chi S\indices{_\mu_\nu}$ \\ 
$M\indices{_\mu_\nu}$ & $M\indices{_\mu_\nu}$ & $M\indices{_\mu_\nu}$ &  $\chi S\indices{_\mu_\nu}$ & $S\indices{_\mu_\nu}$ 
& $\mathbbm{1}, \chi$, ${C}\indices{_\mu_\nu_\rho_\sigma}$ & $M\indices{_\mu_\nu}$\\ 
$C\indices{_\mu_\nu_\rho_\sigma}$ & $C\indices{_\mu_\nu_\rho_\sigma}$ & $C\indices{_\mu_\nu_\rho_\sigma}$ 
& $S\indices{_\mu_\nu}$ & $\chi S\indices{_\mu_\nu}$ & $M\indices{_\mu_\nu}$ & $\mathbbm{1}, \chi, C\indices{_\mu_\nu_\rho_\sigma}  $ \\ 
\end{tabular} \end{ruledtabular}
\end{table*}

\subsection{Lorentz structure of the \texorpdfstring{$(3/2,0) \oplus (0,3/2)$}{(3/2,0)+(0,3/2)}. }

As a final explicit example let us now consider the $j=\frac{3}{2}$ case. The angular momentum matrices are given by
\begin{equation}\begin{split}
\tau_1 &= \frac{1}{2}\begin{pmatrix} 
0 & \sqrt{3} & 0 & 0 \\ \sqrt{3} & 0 & {2} & 0 \\ 0 & {2} & 0 & \sqrt{3} \\  0 & 0 & \sqrt{3} & 0 
\end{pmatrix} \\ 
\tau_2 &= \frac{i}{2} \begin{pmatrix} 
0 & -\sqrt{3} & 0 & 0 \\ \sqrt{3} & 0 & -2 & 0 \\ 0 & {2} & 0 & -\sqrt{3} \\  0 & 0 & \sqrt{3} & 0 
\end{pmatrix}  \\ 
\tau_3 &= \frac{1}{2}\begin{pmatrix}
 3 & 0 & 0 & 0 \\ 0 & 1 & 0 & 0 \\  0 & 0 & -1 & 0 \\ 0 & 0 & 0 & -3 \end{pmatrix}.
\end{split}\end{equation}
 
In this case the external product of states in the basis decomposes as
\begin{equation}\label{eq:j32decom}
\begin{split}
{} &\left(0,0\right)_2 \oplus \left(\frac{3}{2},\frac{3}{2}\right)_2  \oplus(1,0) \oplus (0,1) \\ 
& \oplus(2,0)\oplus(0,2) \oplus(3,0)\oplus(0,3).
\end{split}\end{equation}
Here, besides the $\mathbbm{1}$ and $\chi$ scalars, we have the decomposition
\begin{equation}
2\ \ydiagram{3}  \oplus \ydiagram{1,1} \oplus \ydiagram{2,2} \oplus \ydiagram{3,3}. 
\end{equation}
This corresponds to a pair of third rank totally symmetrical tensors, and the operators
\begin{equation}
\{ M\indices{_\mu_\nu}, C\indices{_\mu_\nu _\rho _\sigma}, D_{\mu\nu \rho \sigma \alpha\beta} \}.
\end{equation}
The $C$ tensor is given by Eq. (\ref{Ctensor}) with the appropriate changes in the $M\indices{_\mu_\nu}$ operators. 
In the calculation of the remaining tensors it is useful to define the quantity
\begin{equation*}
\{ A,B,C \} = A B C + A C B + B A C + B C A +  C A B +  C B A.
\end{equation*}
The symmetric tensor is constructed along the lines of the $j=1$ case. We just quote the final result
\begin{equation}\begin{split} 
S\indices{_\mu_\nu_\rho} = {} &  \ \frac{1}{2} \Pi (- \eta\indices{_0 _\mu} \eta\indices{_0 _\nu} \eta\indices{_0 _\rho}   
+ \eta\indices{_\mu _\nu} \eta\indices{_0 _\rho} + \eta\indices{_\mu _\rho} \eta\indices{_0 _\mu} 
+ \eta\indices{_\rho _\nu} \eta\indices{_0 _\mu}) \\ & 
+  \frac{i}{9} \Pi \left[  7 \eta\indices{_\mu_\nu}  M\indices{_0 _\rho} + 7 \eta\indices{_\mu_\rho} M\indices{_0 _\nu} 
+ 7 \eta\indices{_\nu_\rho} M\indices{_0 _\mu} \right] \\ & 
-  i \Pi \left( \eta\indices{_0 _\mu} \eta\indices{_0 _\nu}  M\indices{_0 _\rho} 
+ \eta\indices{_0 _\mu} \eta\indices{_0 _\rho} M\indices{_0 _\nu}  
+ \eta\indices{_0 _\nu} \eta\indices{_0 _\rho}  M\indices{_0 _\mu} \right) \\ &
+ \frac{2i}{9} \Pi\{ M\indices{_0 _\mu}, M\indices{_0 _\nu},M\indices{_0 _\rho}\} . 
\end{split}\end{equation}To build the sixth-order tensor $D\indices{_\mu_\nu _\rho _\sigma _\alpha_\beta}$, we apply the 
Young projector\ $\begin{ytableau} \mu&\alpha & \rho \\ \nu&\beta &\sigma\\  \end{ytableau}$ to the product 
$M\indices{_\mu_\nu} M\indices{_\rho _\sigma} M\indices{_\alpha_\beta}$ to get
\begin{equation}\begin{split}
 Y\indices{_\mu _\nu  _\rho _\sigma  _\alpha _\beta} &=
  \frac{4}{18} \{ M_{\mu \nu}, M_{\rho \sigma} , M_{\alpha \beta}\} 
+ \frac{2}{18} \{ M_{\mu \nu}, M_{\rho \alpha} , M_{\sigma \beta}\} \\ & 
- \frac{2}{18} \{ M_{\mu \nu}, M_{\rho \beta} , M_{\alpha \sigma}\}
+ \frac{2}{18} \{ M_{\mu \alpha}, M_{\rho \sigma} , M_{\nu \beta}\} \\ & 
- \frac{2}{18} \{ M_{\mu \beta}, M_{\rho \beta} , M_{\nu \alpha}\} 
+ \frac{2}{18} \{ M_{\mu \rho}, M_{\nu \sigma} , M_{\alpha \beta}\} \\ &
- \frac{2}{18} \{ M_{\mu \sigma}, M_{\nu \rho} , M_{\alpha \beta}\} 
+ \frac{1}{18} \{ M_{\mu \rho}, M_{\nu \alpha}, M_{\sigma \beta}\} \\ & 
- \frac{1}{18} \{ M_{\mu \rho}, M_{\nu \beta}, M_{\sigma \alpha}\} 
+ \frac{1}{18} \{ M_{\mu \sigma}, M_{\nu \beta}, M_{\rho \alpha}\} \\ &
- \frac{1}{18} \{ M_{\mu \sigma}, M_{\nu \alpha}, M_{\rho \beta}\} 
+ \frac{1}{18} \{ M_{\mu \alpha}, M_{\nu \sigma}, M_{\rho \beta}\} \\ &
- \frac{1}{18} \{ M_{\mu \alpha}, M_{\nu \rho}, M_{\sigma \beta}\}
+ \frac{1}{18} \{ M_{\mu \beta}, M_{\nu \rho}, M_{\sigma \alpha}\} \\ &
- \frac{1}{18} \{ M_{\mu \beta}, M_{\nu \sigma}, M_{\rho \alpha}\}.
\end{split}\end{equation}

As with the $C$ tensor, this tensor is not traceless and we need to remove the Young-projected contractions, which are of the form
\begin{equation} \begin{split}
Y\indices{_\mu ^\rho _\rho _\sigma  _\alpha _\beta}  = {} & 
- \eta\indices{_\mu _\sigma} M\indices{_\alpha _\beta} 
+ \frac{1}{2} \eta\indices{_\mu _\alpha} M\indices{_\beta _\sigma} 
- \frac{1}{2} \eta\indices{_\mu _\beta} M\indices{_\alpha _\sigma} \\ &
+ \frac{1}{2} \eta\indices{_\sigma _\beta} M\indices{_\mu _\alpha} 
- \frac{1}{2} \eta\indices{_\sigma _\alpha} M\indices{_\mu _\beta} \equiv Y\indices{_\mu _\sigma  _\alpha _\beta}.
\end{split} \end{equation}
The Young-projected trace is proportional to the following tensor
\begin{equation}\begin{split}
\Lambda \indices{_\mu _\nu  _\rho _\sigma  _\alpha _\beta}  = {} & 
\ \eta\indices{_\mu _\rho} Y\indices{_\nu _\sigma  _\alpha _\beta} 
- \eta\indices{_\nu _\rho} Y\indices{_\mu _\sigma  _\alpha _\beta}
- \eta\indices{_\mu _\sigma} Y\indices{_\nu _\rho  _\alpha _\beta}\\ &
+ \eta\indices{_\nu _\sigma} Y\indices{_\mu _\rho  _\alpha _\beta}
+ \eta\indices{_\mu _\alpha} Y\indices{_\nu _\beta  _\rho _\sigma} 
- \eta\indices{_\nu _\alpha} Y\indices{_\mu _\beta  _\rho _\sigma}\\ & 
- \eta\indices{_\mu _\beta} Y\indices{_\nu _\alpha  _\rho _\sigma} 
+ \eta\indices{_\nu _\beta} Y\indices{_\mu _\alpha  _\rho _\sigma}
+ \eta\indices{_\rho _\alpha} Y\indices{_\sigma _\beta  _\mu _\nu}\\ &
- \eta\indices{_\sigma _\alpha} Y\indices{_\rho _\beta  _\mu _\nu}
- \eta\indices{_\rho _\beta} Y\indices{_\sigma _\alpha  _\mu _\nu}
+ \eta\indices{_\sigma _\beta} Y\indices{_\rho _\alpha  _\mu _\nu}.
\end{split}\end{equation}
The final result for the $D$ tensor is 
\begin{equation}
 D\indices{_\mu _\nu  _\rho _\sigma  _\alpha _\beta}= Y\indices{_\mu _\nu  _\rho _\sigma  _\alpha _\beta} 
 + \frac{41}{60} \Lambda\indices{_\mu _\nu  _\rho _\sigma  _\alpha _\beta}.
\end{equation}
This tensor is antisymmetric in the $\mu\nu$, $\rho\sigma$ and $\alpha\beta$ indices, completely symmetric under the 
exchange of these pairs, it is traceless (the contraction of any pair of indices vanishes) and satisfy the generalized 
Bianchi identity
\begin{equation}
D\indices{_\mu _\nu  _\rho _\sigma  _\alpha _\beta} + D\indices{_\mu _\nu_\rho_\alpha _\beta_\sigma }
+ D\indices{_\mu _\nu  _\rho _\beta_\sigma  _\alpha } =0.
\end{equation}

Notice that the $D$ tensor is like a symmetric third-rank tensor in six dimensions (the six possible values of 
the antisymmetric pairs of indices), which has $56$ components. The traceless condition then removes $36$ of those 
components, and the generalized Bianchi identity removes another $6$, leaving only 14 independent components 
which are the degrees of freedom for the $(3,0)\oplus(0,3)$ representation.   

Concerning the \poi\ projector formalism for the $(3/2,0)\oplus(0,3/2)$ representation, the antisymmetric part 
of the space-time tensor $T\indices{_\mu_\nu}$ in Eq.(\ref{pp}) must be constructed with the elements of the basis. 
Here, in principle we can have contractions of $C$ and $D$ tensors among themselves and with products of the metric 
tensor or of the generators. All these possible products are operators that can be expanded in terms of the basis and 
the result must be a rank 2 tensor antisymmetric under the exchange $\mu \leftrightarrow \nu$. Since the only rank 2 
tensor with this property in the basis is the Lorentz generator tensor, these products must be proportional to 
$M\indices{_\mu_\nu}$, thus electromagnetic properties are in this case also of the form 
$g~M\indices{_\mu_\nu}F\indices{^\mu^\nu}$ when we use the gauge principle.

\subsection{The general structure of \texorpdfstring{$(j,0) \oplus (0,j)$}{(j,0)+(0,j)}\ fields.}
In general, for the $(4j+2)$-dimensional representations $(j,0)\oplus(0,j)$ the external product of the states in the basis 
has the decomposition
\begin{equation} 
\left[(j,0) \oplus (0,j)\right]^2 = \bigoplus_{i=0}^{2j}[(i,0)\oplus(0,i)] \oplus 2(j,j) .
\end{equation}
We can  construct, for every $j$, a set of operators which form a basis for this square space. In general this set will contain 
the scalars $\{\mathbbm{1}, \chi\}$, and a pair of symmetrical tensors transforming as $(j,j)$, The chirality and parity 
operators are given by Eqs. (\ref{chi},\ref{Pi}). Parity turns out to be the time component of the totally symmetric 
tensor $S\indices{^{\mu_{1}}^{\mu_{2}}^{...\mu_{2j}}}$ transforming as $(j,j)$. In general chirality and parity 
anticommutes which in turn causes that chirality and the symmetric tensor $S$ also anticommute. The second symmetric 
tensor is given in general as $\chi S\indices{^{\mu_{1}}^{\mu_{2}}^{...\mu_{2j}}}$. In addition the covariant basis contains 
the series 
\begin{equation}
\bigoplus_{i=1}^{2j}[(i,0)\oplus(0,i)] =  \ydiagram{1,1} \oplus \ydiagram{2,2} \oplus \ydiagram{3,3} \oplus \ldots 
\end{equation}
which gives the generators, plus a series of generalizations of the Weyl tensor
\begin{equation}
\{M\indices{_\mu_\nu}, C\indices{_\mu_\nu _\rho _\sigma}, D\indices{_\mu_\nu _\rho _\sigma _\alpha _\beta}, 
E\indices{_\mu_\nu _\rho _\sigma _\alpha _\beta _\tau _\delta} ,\ldots\}.
\end{equation}
These tensors are to be constructed by taking the product of $2j$ generators, applying the Young 
projector and removing all contractions. 

Concerning the \poi\ projector formalism for the $(j,0)\oplus(0,j)$, the antisymmetric part of the space-time tensor 
$T\indices{_\mu_\nu}$ in Eq.(\ref{pp}) in general must be constructed with the elements of a basis of this space. 
When using our covariant basis, due to the properties of the $C$, $D$, $E...$ tensors coming from the 
Young projectors it is clear that the result found for $j=1/2, 1, 3/2$ is valid for any $j$. The contractions of these tensors 
among themselves, with the metric tensor or with the generators yielding a rank $2$ antisymmetric tensor is necessarily 
proportional to $M\indices{_\mu_\nu}$ or it vanishes. In consequence, for arbitrary $j$, the most general space-time tensor 
in Eq.(\ref{pp}) is given by
\begin{equation}
T_{\mu\nu}=\eta_{\mu\nu} -i g M_{\mu\nu}.
\label{Tmunu}
\end{equation}
There are two direct consequences of this result: i) the multipole electromagnetic moments of a spin $j$ particle in this 
formalism are dictated by two free parameters, the electric charge $e$ and the gyromagnetic factor $g$, and ii) the propagation 
of spin $j$ waves in an electromagnetic background is causal. In the next section we elaborate on these points. 

\section{Electromagnetic structure and causal propagation for the \texorpdfstring{$(j,0) \oplus (0,j)$}{(j,0)+(0,j)}\ representation.}

In the \poi\ projector formalism, the Lagrangian for an interacting elementary particle transforming in the 
$(j,0)\oplus(0,j)$ representation is
\begin{equation}
\mathcal{L}=\overline{D^{\mu}\psi}T\indices{_\mu_\nu}D^{\nu}\psi-m^{2}\overline{\psi}\psi,
\end{equation}
where $\overline{\psi} = \psi^{\dagger}\Pi$, $D^{\mu} = \partial^{\mu} + i e A^{\mu}$ for a particle of charge 
$e$ and $T\indices{_\mu_\nu}$ stands for a space-time tensor. The \poi\ projector fixes only the symmetric part 
of this tensor. The anti-symmetric part must be constructed in terms of the basis for the operators acting on the 
$(j,0)\oplus(0,j)$ representation. According to results in the last section, the most general space-time tensor is 
given by the tensor in Eq. \ref{Tmunu}. This is the tensor used in \cite{DelgadoAcosta:2012yc} to calculate the 
multipole moments of a particle transforming in these representations for $j=1/2,1,3/2$. It  is shown there that 
the multipole moments are dictated solely by the two parameters appearing in the Lagrangian, the charge $e$ and the 
gyromagnetic factor $g$. Our results in the previous sections put these calculations on a firm basis and allow us 
to generalize them to particles of arbitrary spin $j$. 

The electromagnetic current is simply calculated to be
\begin{equation}\begin{split}
 J_{\mu}(p^{\prime},\lambda^{\prime};p,\lambda) = {} & e\overline{u}(p^{\prime},\lambda^{\prime}) \left[ \left( p^{\prime}+p\right)_{\mu} 
\right. \\ & \left.
+ i g M\indices{_\mu_\nu}(p^{\prime}-p)^{\nu}\right]  u(p,\lambda).
\end{split}\end{equation}
The multipole moments of this current can be calculated from the charge and current densities using the Breit frame where
\begin{equation}
p^{\prime}=({\omega}/{2},{\mathbf{q}}/{2}),\quad p=({\omega}/{2},-{\mathbf{q}}/{2}).
\end{equation}
In terms of the Breit current defined by
\begin{equation}
J_{\mu}^{B}(\mathbf{q},j,\lambda)=\frac{1}{\omega}J_{\mu}(\mathbf{p}^{\prime},\lambda;\mathbf{p},\lambda),
\label{BJ}
\end{equation}
with $\omega=\sqrt{4m^{2}+\mathbf{q}^{2}}$, the electromagnetic moments for a particle of spin $j$ and polarization 
$\lambda$ are given by
\begin{align}
Q_{E}^{l}(\mathbf{q},j,\lambda)  &  =\left.  b^{l0}(-i\nabla_{\mathbf{q}%
})\varrho_{E}(\mathbf{q},j,\lambda)\right\vert _{\mathbf{q}=0},\nonumber\\
Q_{M}^{l}(\mathbf{q},s,\lambda)  &  =\frac{1}{l+1}\left.  b^{l0}%
(-i\nabla_{\mathbf{q}}){\varrho_{M}}(\mathbf{q},j,\lambda)\right\vert
_{\mathbf{q}=0}, \label{qmformulas}%
\end{align}
where the electric, $\varrho_{E}(\mathbf{q},j,\lambda)$, and the magnetic, $\varrho_{M}(\mathbf{q},j,\lambda)$, 
densities are given by
\begin{equation}\begin{split}
\varrho_{E}(\mathbf{q},j,\lambda) &= J_{B}^{0}(\mathbf{q},j,\lambda
), \\ 
\varrho_{M}(\mathbf{q},j,\lambda) &= \nabla_{\mathbf{q}}\cdot
\lbrack\mathbf{J}^{B}(\mathbf{q},j,\lambda)\times\mathbf{q}],
\end{split}\end {equation}
and the $b^{l0}$ operators are given by
\begin{equation}
b^{l0}(\mathbf{r})=l!\sqrt{{4\pi}/(2l+1)}r^{l}Y_{l0}(\Omega).
\end{equation}
Explicitly, for $l=1,2,...8$ these operators read
\begin{equation}\begin{split}\label{bl0}
b^{00}(\mathbf{r})   = {} & 1,  \\
b^{10}(\mathbf{r})   = {} & z, \\
b^{20}(\mathbf{r})   = {} & 3z^{2} - r^{2}, \\
b^{30}(\mathbf{r})   = {} & 3z\left(  5z^{2} - 3r^{2}\right),  \\
b^{40}(\mathbf{r})   = {} & 3\left(  35z^{4} - 30z^{2}r^{2} + 3r^{4}\right), \\
b^{50}(\mathbf{r})   = {} & 15z(63z^{4} - 70z^{2}r^{2} + 15r^{4}), \\
b^{60}(\mathbf{r})   = {} & 45(231z^{6}-315z^{4}r^{2}+105z^{2}r^{4}-5r^{6}), \\
b^{70}(\mathbf{r})   = {} & 315z(429z^{6} - 693z^{4}r^{2} + 315z^{2}r^{4} - 35r^{6}), \\
b^{80}(\mathbf{r})   = {} & 315(6435z^{8} - 12012z^{6}r^{2} + 6930z^{4}r^{4} \\ & - 1260z^{2}r^{6} + 35r^{8}). 
\end{split}\end{equation}
The electromagnetic current can be rewritten as
\begin{equation}\begin{split}
J_{\mu}(p^{\prime},\lambda^{\prime};p,\lambda) = {} & e\overline{u}(0,\lambda^{\prime}) 
\left[ \left( p^{\prime} + p\right)_{\mu} B(-p^{\prime}) B(p) \right. \\ & \left.
+ i g B(-p^{\prime}) M\indices{_\mu_\nu} B(p)(p^{\prime}-p)^{\nu} \right]u(0,\lambda),
\end{split}\end{equation}
where $B(p)$ stands for the boost operator.

In the Breit frame we obtain
\begin{equation}
B(-p^{\prime})=\exp\left[  i\mathbf{K} \cdot \boldsymbol\varphi^{\prime} \right] ,\qquad B(p)
=\exp\left[ -i \mathbf{K} \cdot \boldsymbol\varphi \right]
\end{equation}
with%
\begin{equation}
\cosh\varphi^{\prime}=\frac{\omega}{2m}=\cosh\varphi,\qquad\sinh\varphi^{\prime} = \frac{\mathbf{|q}|}{2m} = \sinh\varphi.
\end{equation}
In terms of the unitary vector $\mathbf{n}=\mathbf{q/|q}|$ the corresponding angles are
\begin{equation}
\boldsymbol\varphi = -\mathbf{n}\varphi,\qquad \boldsymbol\varphi^{\prime}=\mathbf{n}\varphi,
\end{equation}
hence
\begin{equation}
B(-p^{\prime})=B(p)=\exp\left[  i\mathbf{K}\cdot\mathbf{n}\varphi\right]  \equiv B(\mathbf{q}).
\end{equation}
The time component of the electromagnetic current then becomes
\begin{equation}\begin{split}
J_{0}(p^{\prime},\lambda^{\prime};p,\lambda) = {} & e\overline{u}(0,\lambda^{\prime}) 
\left[  \left(  \omega - i g\mathbf{K}\cdot\mathbf{n|q}|\right) \right. \\ & \left. \times  
\exp\left[ i 2\mathbf{K} \cdot \mathbf{n}\varphi\right]  \right]  u(0,\lambda),
\label{rhoe}
\end{split}\end{equation}

For the representations $\left(  j,0\right)\oplus \left( 0,j\right) $ the rotations and boosts generators are 
related as $i \mathbf{K} = -\chi\mathbf{J}$, which when used for the charge density 
in the Breit frame yield
\begin{equation}
\varrho_{E}(\mathbf{q},j,\lambda)=e\overline{u}(0,\lambda^{\prime}%
)O(\rho,x)u(0,\lambda),
\end{equation}
where $x=\frac{|\mathbf{q|}}{2m}$, $\rho=\mathbf{J}\cdot\mathbf{n}$ and the operator $O(\rho,x)$ is given by%
\begin{equation}
O(\rho,x)=\left(  1+\frac{g\rho\chi x}{\sqrt{1+x^{2}}}\right)  \exp\left[
-2\chi\mathbf{\rho}\sin h^{-1}\left(  x\right)  \right]  . \label{Eoperator}%
\end{equation}
The electric multipoles involve the matrix elements of derivatives with respect to $q_{i}$ of this operator; 
for this reason, it is convenient to expand $O(\rho,x)$ in powers of $x$. Expanding and using $\chi^2=1$ we get
\begin{equation}\begin{split}
O(\rho,x) = {} & 1 + (g - 2) \chi \rho x - 2(g-1)\rho^{2}x^{2} \\ & + \frac{\rho}{6}\left[ \left( 12g - 8\right)  \rho^{2} 
- \left( 3g - 2\right)  \right] \chi x^{3}+...
\end{split}\end{equation}
The calculation of the $l$-th multipole requires the $l$-th derivatives of this operator, with only the order 
$x^{l}$ term contributing. Using $\{\chi,\Pi\}=0$, $[\chi,\mathbf{J}]=0$ and $[\Pi,\mathbf{J}]=0$, it can be shown that 
the matrix elements of odd powers of $x$ between states of the same parity vanish. As a consequence, odd electric multipole moments vanish for particles 
of well defined parity. Skipping odd terms in the expansion we rewrite the operator in Eq. (\ref{Eoperator}) up to order 
$x^{8}$ as
\begin{equation}\begin{split}
O(\rho,x)  = {} & 1 - 2(g-1) \rho^{2} x^{2} - \frac{2 \left(  2g-1\right)}{3}\left[ \rho^{4} - \rho^{2} \right]  x^{4} \\ & 
- \frac{4 \left(3 g -1 \right)}{45}\left[ \rho^{6} - 5\rho^{4} + 4 \rho^{2}\right] x^{6} \\
&  - \frac{2 (4g - 1)}{315}\left[ \rho^{8} -14 \rho^{6} + 49\rho^{4} - 36 \rho^{2} \right] x^{8} \\ &+ \ldots 
\end{split}\end{equation}
The calculation of the electric multipole moments for arbitrary values of $j$ and $\lambda$ is now straightforward. The first five 
non--vanishing electric multipole moments, for arbitrary $j$, are given by%
\begin{widetext}\begin{equation}\begin{split} \label{QEl}
Q_{E}^{0}(j,\lambda)  = {} & e    \\
Q_{E}^{2}(j,\lambda)  = {} & -\frac{e(g-1)}{m^{2}} \langle \mathbf{J}^{2} - 3J_{z}^{2} \rangle \\
Q_{E}^{4}(j,\lambda)  = {} & -\frac{e}{m^{4}} 3(2g-1) \langle 3\mathbf{J}^{4} - 30\mathbf{J}^{2}J_{z}^{2} 
+ 35J_{z}^{4} - 6\mathbf{J}^{2}+25J_{z}^{2}\rangle \\
Q_{E}^{6}(j,\lambda)  = {} & -\frac{e}{m^{6}}(3g-1) 45\langle 5\mathbf{J}^{6} - 105\mathbf{J}^{4}J_{z}^{2} 
+ 315\mathbf{J}^{2}J_{z}^{4} - 231J_{z}^{6} - 40\mathbf{J}^{4}   + 525\mathbf{J}^{2}J_{z}^{2}  - 735J_{z}^{4} 
+ 60\mathbf{J}^{2} - 294 J_{z}^{2}\rangle  \\
Q_{E}^{8}(j,\lambda)  = {} & -\frac{e}{m^{8}}(4g-1) 315 \langle 35 J^8-1260 J^6 Jz^2 - 700 J^6 + 6930 J^4 Jz^4 
+ 18270 J^4 Jz^2 + 3780 J^4 - 12012 J^2 Jz^6 6  \\ & 
- 64680 J^2 Jz^4 - 59388 J^2 Jz^2 - 5040 J^2 + 6435 Jz^8 + 54054 Jz^6 + 93555 Jz^4 + 27396 Jz^2 \rangle
\end{split}\end{equation}\end{widetext}
where we used the shorthand notation
\begin{equation}
 \langle O\rangle \equiv \overline{u}(0,\lambda) O u(0,\lambda).
\end{equation}
It is worth remarking that, for a given $j$, the special combination of $\mathbf{J}^{2}$ and $J_{z}$ appearing in 
$Q_{E}^{l}(j,\lambda)$ in Eqs. (\ref{QEl}) vanishes for $l>2j$. This is consequence of the full algebraic structure 
of the covariant basis for $(j,0)\oplus(0,j)$ representation, which at this level manifests in the fact that the 
rotation generators satisfy
\begin{equation}
\prod\limits_{\lambda=-j}^{j}
\left(  \mathbf{J}\cdot\mathbf{n-\lambda}\right)  =0,
\end{equation}
for an arbitrary unitary vector $\mathbf{n}$. For $\mathbf{n=k}$ this relation lowers the powers of $J_{z}$ 
appearing in $Q_{E}^{l}(j, \lambda)$ and causes it to vanish for $l>2j$. The simplest example is $j=1/2$ in 
whose case $J_{z}^{2}=1/4$. In this case, the combinations of $\mathbf{J}^{2}$ and $J_{z}$ appearing in 
$Q_{E}^{l}(j,\lambda)$ for $l>1$ reduces to the unity operator ($\mathbf{J}^{2}$ is diagonal) with vanishing 
coefficient vanishes,  as can be easily checked. For $j=1$ we get $J_{z}^{3}=J_{z}$ in which case the 
combinations of $\mathbf{J}^{2}$ and $J_{z}$ appearing in $Q_{E}^{l}(j,\lambda)$ for $l>2$ reduces to a linear 
combination of the unity operator and $J_{z}$ with vanishing coefficients. Similar results are obtained for 
higher values of $j$. Therefore, we understand the well known fact that a spin $j$ particle can have at 
most $2j$ non-vanishing electric multipole moments as a consequence of the full algebra satisfied by the 
elements of the basis of operators acting on the $(j,0)\oplus(0,j)$ representation.

As for the magnetic current a similar calculation yields%
\begin{equation}
\varrho_{M}(\mathbf{q},j,\lambda)=i e g[\overline{u}(0,\lambda)\left[
\nabla_{\mathbf{q}}\cdot\mathbf{M(q},j)\right]  u(0,\lambda)]
\end{equation}
with%
\begin{equation}
\mathbf{M(q},j)=\frac{1}{\omega}B(\mathbf{q})\left[  \left(  \mathbf{J}%
\cdot\mathbf{q}\right)  \mathbf{q-|q|}^{2}\mathbf{J}\right]  B(\mathbf{q}).
\label{Mqj}
\end{equation}
To evaluate the magnetic multipoles we calculate the derivatives of this operator and then evaluate them at 
$\mathbf{q}=0$. Notice that the $l$-th magnetic moment receives contributions only of the term $\mathbf{q}^{l+1}$ 
in the expansion of $\mathbf{M(q},j)$. In particular, the calculation of $Q_{M}^{1}$ does not involve the specific 
structure of the boost operator. For $l$ even, the term $\mathbf{q}^{l+1}$ in the expansion of $\mathbf{M(q},j)$ 
contains a factor $\chi$ and the even magnetic multipole moments vanish for the same reason as the odd electric 
moments do. The non-vanishing lowest order magnetic multipoles are given by
\begin{widetext}\begin{equation}\begin{split} \label{QMl}
Q_{M}^{1}(j,\lambda)  = {} & \frac{e g}{2m}\langle J_{z}\rangle  \\
Q_{M}^{3}(j,\lambda)  = {} & \frac{e g}{2m^{3}} 9\langle 
3\mathbf{J}^{2}J_{z}-5J_{z}^{3}-J_{z}\rangle \\
Q_{M}^{5}(j,\lambda)  = {} & \frac{e g}{2m^{5}} \frac{75}{2} \langle 
15\mathbf{J}^{4}J_{z} -70 \mathbf{J}^{2}J_{z}^{3} + 63 J_{z}^{5} 
- 50 \mathbf{J}^{2}J_{z} + 105 J_{z}^{3} + 12 J_{z}\rangle. \\
Q_{M}^{7}(j,\lambda)  = {} & \frac{e g}{2m^{7}} \frac{2205}{2} \langle 
35\mathbf{J}^{6}J_{z} -315 \mathbf{J}^{4}J_{z}^{3}  +693 \mathbf{J}^{2}J_{z}^{5} -429 J_{z}^{7} 
-385 \mathbf{J}^{4}J_{z} \\ & + 2205 \mathbf{J}^{2}J_{z}^{3} - 2310 J_{z}^{5}   + 882 \mathbf{J}^{2}J_{z} - 2121 J_{z}^{3} - 180 J_{z}\rangle. \\
\end{split}\end{equation}\end{widetext}
We would like to remark that, beyond the gauge principle, our calculation depends only on the space-time structure 
of the $(j,0)\oplus(0,j)$ representation: i) the \poi\ projector, ii) the explicit form of the Lorentz generators 
and iii)  the properties of the parity-based covariant basis. This is, therefore, a calculation from first principles, 
and our main result in this section is that, beyond the electric charge, all multipole moments for an elementary 
particle in this formalism are dictated by a single Lorentz structure, namely $M\indices{_\mu_\nu}$, and consequently 
all multipole moments are dictated by the value of the corresponding constant, $g$, the gyromagnetic factor. This is a 
free parameter in the \poi\ projector formalism but for low values of $j$ it has been fixed to be $g=2$ 
\cite{Napsuciale:2006wr, Napsuciale:2007ry, AngelesMartinez:2011nt, DelgadoAcosta:2012yc, VaqueraAraujo:2012qa}. 
On the other hand, there is a variety of consistency arguments for this value to be universal  (see \cite{Holstein:2006wi} 
for a review and further references) and we consider this value in our numerical computations below.

The existence of relations among multipole moments for elementary particles has been noticed before. The calculation of 
$Q^{1}_{M}$, $Q^{2}_{E}$ and $Q^{3}_{M}$ for $j=1/2,1,3/2$ in the \poi\ projector formalism was done in Ref. 
\cite{DelgadoAcosta:2012yc}. Our alternative derivation confirm results in this work but the form of $Q^{3}_{M}$ in 
Eq. (4.52d) of \cite{DelgadoAcosta:2012yc} is valid only for $j=3/2$. The relation between $Q^{3}_{M}$ and the matrix elements of the 
rotation generators valid for every value of $j$ is given in Eqs.(\ref{QMl}) and for $j=3/2$ agrees with Eq. (4.52d) of 
\cite{DelgadoAcosta:2012yc}. 
Recently, a systematic calculation of the multipole moments  for particles of arbitrary spin $j$ transforming in the 
Rarita-Schwinger representations was performed in Ref. \cite{Lorce:2009bs}, using a well motivated ansatz for the
electromagnetic current written in terms of the covariant Lorentz structures and multipole form factors $G_{El}(q^{2})$ 
and $G_{Ml}(q^{2})$. As remarked there, as defined in this reference, $G_{El}(0)$ and $G_{Ml}(0)$ have an interpretation 
of electromagnetic multipole moments only for $l=0,1,2$. For $l>2$ these form factors at zero transferred momentum are 
related to conventional electromagnetic multipole moments through $l$-dependent factors. The multipole form factors 
$G_{El}(0)$ and $G_{Ml}(0)$ calculated in that work, are related to our multipole moments in Eqs. (\ref{QEl},\ref{QMl}) 
with $\lambda=j$. The appropriate quantities to compare with results obtained in Ref \cite{Lorce:2009bs} are 
\footnote{A different normalization factor $1/2m$ is is used in Ref \cite{Lorce:2009bs} for the Breit current instead of 
our $1/\omega$ in Eq. (\ref{BJ}) but this is irrelevant in the calculation of the multipoles.}
\begin{equation}
\widehat{G}_{El}=\frac{m^{l}}{e}\frac{2^{l}}{\left(  l!\right)  ^{2}}Q_{E}%
^{l}(j,j),\quad\widehat{G}_{Ml}=\frac{2m^{l}}{e}\frac{2^{l}}{\left(
l!\right)  ^{2}}Q_{M}^{l}(j,j). 
\label{QElQMlmod}%
\end{equation}

\begin{table}[tbp] \centering 
\caption{ Multipole moments normalized according to Eq. (\ref{QElQMlmod}).} 
\label{tb:multipolemoments}
\begin{ruledtabular} \begin{tabular}[c]{c c c c c c c c c c }
$j$ & $\widehat{G}_{E0}$ & {$\widehat{G}_{M1}$} & {$\widehat{G}_{E2}$} &
{$\widehat{G}_{M3}$} & {$\widehat{G}_{E4}$} & {$\widehat{G}_{M5}$} & {$\widehat{G}_{E6}$} 
& {$\widehat{G}_{M7}$} & {$\widehat{G}_{E8}$}\\ \colrule
$0$   & $1$ & $0$ & $0$  & $0$    & $0$    & $0$   & $0$   & $0$   & $0$  \\ 
$1/2$ & $1$ & $1$ & $0$  & $0$    & $0$    & $0$   & $0$   & $0$   & $0$  \\ 
$1$   & $1$ & $2$ & $1$  & $0$    & $0$    & $0$   & $0$   & $0$   & $0$  \\ 
$3/2$ & $1$ & $3$ & $3$  & $-3$   & $0$    & $0$   & $0$   & $0$   & $0$  \\ 
$2$   & $1$ & $4$ & $6$  & $-12$  & $-3$   & $0$   & $0$   & $0$   & $0$  \\ 
$5/2$ & $1$ & $5$ & $10$ & $-30$  & $-15$  & $5$   & $0$   & $0$   & $0$  \\ 
$3$   & $1$ & $6$ & $15$ & $-60$  & $-45$  & $30$  & $5$   & $0$   & $0$  \\ 
$7/2$ & $1$ & $7$ & $21$ & $-105$ & $-105$ & $105$ & $35$  & $-7$  & $0$  \\ 
$4$   & $1$ & $8$ & $28$ & $-168$ & $-210$ & $280$ & $140$ & $-56$ & $-7$ \\ 
\end{tabular}
\end{ruledtabular}
\end{table}
We list the values for $\widehat{G}_{El}$ and $\widehat{G}_{Ml}$ predicted by the \poi\ projector formalism for the 
lowest values of $\ j$ and $l$ in Table \ref{tb:multipolemoments}. In general, spin-$j$ particles  transforming in 
different representations of the Lorentz group have different electromagnetic properties.  This has been noticed before 
in Ref. \cite{DelgadoAcosta:2012yc} based on a similar calculation for for $j=1, 3/2$.  The only representation-independent 
multipole moments are the charge and the magnetic moment. This is not a surprising result because these are the only 
multipole moments that are independent of the specific structure of the boost operators. The electric charge is completely 
independent of the Lorentz structure (it is associated to a global symmetry), whereas the magnetic moment is related to the 
rotation generators, which for a fixed value of $j$ have the same algebraic structure even if they are embedded in different 
Lorentz representations. Beyond these multipoles, it is clear from Eqs. (\ref{rhoe},\ref{Mqj}) that the specific structure of 
the boost operators becomes important, and since this structure is different for different representations, we should not 
expect the same multipole moments, a fact reflected in Table \ref{tb:multipolemoments} when compared with Table I of 
Ref. \cite{Lorce:2009bs}. In general, the quantities in Eq. (\ref{QElQMlmod}) for the 
$(j,0)\oplus(0,j)$ representation obtained here are related to those for the Rarita-Schwinger representation obtained 
in Ref. \cite{Lorce:2009bs} by 
\begin{equation}
\widehat{G}_{El}=\left(1 - \frac{g l}{2} \right)  G_{El}(0), \quad \widehat{G}_{Ml}=\frac{g l}{2}~G_{Ml}(0).
\end{equation}

Finally, we would like to remark on another important side result of our construction of the parity-based covariant 
basis for operators acting on the $(j,0)\oplus(0,j)$ representation: the restriction of the antisymmetric part of the 
space-time tensor $T\indices{_\mu_\nu}$ to be given solely by the Lorentz generators tensor $M\indices{_\mu_\nu}$ yields 
causal propagation of spin $j$ waves in an electromagnetic background. Indeed, this problem has been studied in 
Ref. \cite{Delgado-Acosta:2013nia} for $j=1$ under the assumption that the most general tensor is given by 
Eq. (\ref{Tmunu}) for $j=1$. Once we have proved that this is indeed the case and that this result is valid for 
arbitrary $j$, the generalization to the propagation of high spin waves is straightforward. The gauged classical equation 
of motion for arbitrary $j$ can be rewritten as 
\begin{equation}
\left(  D^{\mu}D_{\mu} + \frac{g}{2} M\indices{_\mu_\nu} F\indices{^\mu^\nu} + m^{2}\right)  \psi=0.
\label{coupledeq}
\end{equation}
Notice that Eq. (\ref{coupledeq}) is a set of coupled equations for the $2(2j+1)$ components of the spinor $\psi_{i}$. 
Using the form of the generators in Eq. (\ref{LG}) it is easy to see that all the components have 
a second time derivative. The characteristic determinant is the
determinant of the operator $O(n)$ obtained from the highest derivatives in
this equation when derivatives of the field $\partial^{\mu}\psi$ are replaced
by a constant four-vector $n^{\mu}$. The nature of the propagation of classical waves is determined by 
the vanishing of the characteristic determinant
\begin{equation}
\det [O(n)]=\left(  n^{2}\right)  ^{2(2j+1)}=0.
\end{equation}
The solutions for the time-like components of $n^{\mu}$ are 
$n^{0}=\sqrt{n_{x}^{2}+n_{y}^{2}+n_{z}^{2}}$ which are always real and according to the
Currant-Hilbert criterion the propagation of spin $j$ waves in the \poi\ projector formalism is causal.

Since all the components $\psi_{i}$ have a second time derivative either in the free  or in 
the interacting case, as discussed in \cite{Delgado-Acosta:2013nia} for the case $j=1$, this formalism 
actually describes the dynamics of a degenerate parity doublet. As pointed by Weinberg\cite{Weinberg:1995mt}, 
when the field is built as in \ref{eq:fieldD} the properties of the one-particle states described by the field 
are related to the properties of the coefficients $\omega_l(\Gamma),{\omega}^c_l(\Gamma)$ which, beyond mass and 
spin, furnishes a record of the choice made to ensure the desired properties of the field under discrete 
transformations. In the \poi\ projector formalism, the dynamics is dictated solely by the 
projection onto Poincar\'e eigensubspaces and the appropriate degrees of freedom for the single-particle states  
are fixed through a judicious choice of the coefficients $\omega_l(\Gamma),\omega^{c}_l(\Gamma)$. The spacetime 
properties of the parity operator are such that it commutes with the \poi\ projector. This justifies our use of 
definite-parity states for the calculation of the electromagnetic moments.

\section{Conclusions}
In this work we study the transformation properties of the rest-frame parity operator under Lorentz transformations for 
the $(j,0)\oplus (0,j)$ representation spaces. We show that while rotating as a scalar, under boosts its transformation 
properties involve the Jordan algebra of the generators, which is representation dependent. Using these general 
properties and the Young projectors for $\so(1,3)$ we show that the rest-frame parity operator transforms as the 
completely temporal component of a symmetric traceless tensor of rank $2j$. We provide an algorithm for the calculation 
of a covariant basis for arbitrary $j$. For a given $j$, this basis contains the corresponding identity ($\mathbbm{1}$) 
and chirality ($\chi$) operators, the Lorentz generators ($M\indices{_\mu_\nu}$) and two symmetric traceless tensors of rank $2j$. 
The time component of the first symmetric traceless tensor denoted by $S^{\mu_1\mu_2...\mu_{2j}}$ is precisely the 
rest-frame parity operator and the second symmetric traceless operator is given by the product 
$\chi S^{\mu_1\mu_2...\mu_{2j}}$. In addition, for $j>1$ the basis contain 
tensor operators with the symmetry properties of the Weyl tensor and its generalizations. 

We explicitly construct the basis for $j=1/2,1,3/2$. For $j=1/2$ we reproduce the conventional Dirac basis and rest-frame 
parity is the time component of a four-vector operator (in this case the``symmetric'' tensor of rank $2j=1$) that turns out 
to be the conventional Dirac matrices $\gamma^{\mu}$; the chirality operator coincides with the $\gamma^{5}$ Dirac matrix. 
For $j=1$ the basis is given by 
$\left\{ \mathbbm{1},\chi, S\indices{_\mu_\nu},  \chi S\indices{_\mu_\nu}, M\indices{_\mu_\nu}, C\indices{_\mu_\nu _\alpha_\beta} \right\} $. 
We explicitly construct the symmetric traceless tensor $S\indices{_\mu_\nu}$ and the 
$C\indices{_\mu_\nu _\alpha_\beta}$ tensor which has the symmetries of the Weyl tensor. These two tensors involve the Jordan algebra 
of the generators. For $j=3/2$ the basis contains the operators $\left\{ \mathbbm{1},\chi, S\indices{_\mu_\nu_\rho},  \chi S\indices{_\mu_\nu_\rho}, M\indices{_\mu_\nu}, C\indices{_\mu_\nu_\alpha_\beta}, D\indices{_\mu_\nu_\rho_\sigma_\alpha_\beta} \right\} $.  
The $C$ tensor has the same form  as the $j=1$ case just replacing the $j=1$ generators by those of $j=3/2$. We give explicit 
expressions for the  $S\indices{_\mu_\nu_\rho}$ and $D\indices{_\mu_\nu_\rho_\sigma_\alpha_\beta}$ tensors.

The formulation of theories for particles, either elementary or composite, transforming in these representation can be 
done using our covariant basis which has a clear physical interpretation in terms of parity or chirality properties. 
In particular, in the \poi\ projector formalism for the $(j,0)\oplus (0,j)$ representations we find that the antisymmetric 
part of the involved space-time tensor is given by $M_{\mu \nu}$ for all $j$. This simple  structure yields has two direct 
physical consequences: i) the multipole moments of elementary spin $j$ particles are not independent and dictated by the value of the 
gyromagnetic factor $g$, ii) the propagation of spin $j$ waves in an electromagnetic background in causal. 

We calculate the multipole moments and compare with existing calculations in the literature. We conclude that except for 
$Q^{0}_{E}$ and $Q^{1}_{M}$ the multipole moments are representation specific. The universality of $Q^{0}_{E}$ is due to a 
global symmetry while that of $Q^{1}_{M}$ is due to the fact that it involves only the algebraic properties of the generators 
of rotations which are independent of the chosen Lorentz  representation. Higher multipole moments depend on the algebraic 
structure of the boost generators which is different for different Lorentz representations. This difference in the complete 
algebraic structure is at the root of the representation dependence of higher multipole moments.

\begin{center}
{\bf Acknowledgments}
\end{center}
Work supported by CONACyT under project \# 156618 and by DAIP-UG.

\bibliography{highspin}

\end{document}